# Synthesis, Growth Mechanism, and Photocatalytic Properties of Metallic-Bi/$Bi_{13}S_{18}Br_2$ Nano-Bell Heterostructures

Anna Cabona[1,2], Stefano Toso[1,3], Andrea Griesi[4], Martina Rizzo[1,2], Michele Ferri[1], Pascal Rusch[1], Giorgio Divitini[4], Julia Perez-Prieto*[5], Raquel E. Galian*[5], Ilka Kriegel*[2], Liberato Manna*[1]

[1] Nanochemistry Department, Italian Institute of Technology, Via Morego 30, 16163 Genova, Italy
[2] Department of Applied Science and Technology, Politecnico di Torino, Corso Duca degli Abruzzi 34, 10129 Turin, Italy
[3] Lund University, Division of Chemical Physics, Naturvetarvägen 14, 221 00 Lund, Sweden
[4] Electron Spectroscopy and Nanoscopy, Italian Institute of Technology, Via Morego 30, 16163 Genoa, Italy
[5] Institute of Molecular Science, University of Valencia, c/Catedrático José Beltrán Martínez 2, 46980 Paterna, Valencia, Spain

**TOC**

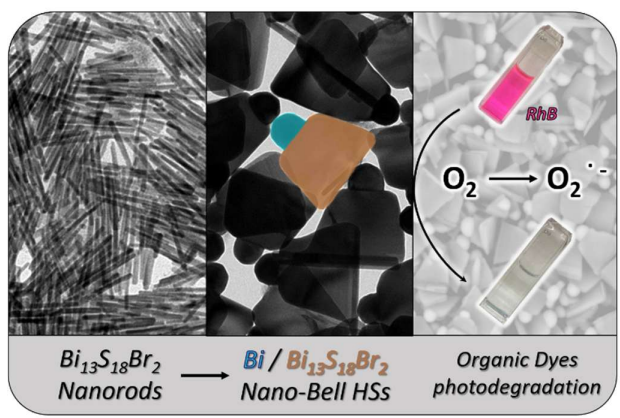

**ABSTRACT**

We report the synthesis of bell-shaped Bi/$Bi_{13}S_{18}Br_2$ metal/semiconductor heterostructures as a photocatalyst based on non-toxic and Earth-abundant elements. Their unique morphology arises from a multi-step growth process, involving 1) the nucleation of $Bi_{13}S_{18}Br_2$ nanorods, 2) the reduction of a metallic-Bi domain on their surface induced by *N,N*-didodecylmethylamine, and 3) the heterostructure accretion by a localized reaction at the Bi/$Bi_{13}S_{18}Br_2$ interface promoted by Ostwald ripening. These heterostructures display remarkable stability in polar solvents, remaining almost unaffected by prolonged exposure to isopropanol and water, and exhibit high photocatalytic efficiency for the degradation of organic dyes (i.e., Rhodamine-B and Methylene Blue) under visible-light irradiation, with good recyclability. Additionally, preliminary tests demonstrate $CO_2$ reduction capabilities, which make them promising for both the photocatalytic degradation of pollutants and photo-electro $CO_2$ conversion. The straightforward synthesis process and the use of non-toxic and earth-abundant elements offers significant potential for sustainable energy conversion technologies.

Colloidal nano-heterostructures enable combining the properties of different materials into a single functional nanocomposite, and may lead to the emergence of new properties stemming from the interaction of the two materials.[1,2,3,4] Among the various heterostructures reported to date, those composed of a metal and a semiconductor stand out as promising candidates for light harvesting applications, as the semiconductor domain can convert light into electron-hole pairs and the metal domain makes them easy to access. Relevant examples include Au-CdSe,[5] Pt-CdS[6] and M-CsPbBr$_3$ (M = Pd,Pt) heterostructures,[7] which exhibit promising performance as light-harvesters,[8,9,10] photodetectors,[11] and photocatalysts.[12,13,14]

A class of semiconductors recently considered in the preparation of colloidal heterostructures are metal chalcohalides,[15,16,17] which can be synthesized using methods and chemicals similar to those employed for metal halides (e.g., CsPbBr$_3$) but do not suffer from the intrinsic lability of metal halides.[18,19] Among them, bismuth-based chalcohalides[20,21,22] are noteworthy due to their low cost and low toxicity, promising photovoltaic[23] and thermoelectric[24,25,26] properties, and well-established solid-state chemistry.[27,28] Colloidal synthesis routes have been proposed for two bismuth chalcohalide phases: the orthorhombic BiSX (X = Cl, Br, I)[29] and the hexagonal Bi$_{13}$S$_{18}$X$_2$ (X = Br, I).[30] In parallel, reports on CsPbBr$_3$/PbBi$_2$S$_4$ heterostructures have shown how semiconductors containing bismuth and sulfur can be integrated in colloidal heterostructures.[31,32]

In this work, we demonstrate the growth of metallic-Bi/Bi$_{13}$S$_{18}$Br$_2$-chalcohalide heterostructures exhibiting a bell-shape (Figure 1a). This was made possible by introducing substantial modifications in the synthesis procedure for bismuth chalcohalide nanocrystals, originally developed by Quarta *et al.* in collaboration with some of us.[33] In that work, nanocrystals were grown by reacting a Bi-oleate solution with different amounts of sulfur and halide precursors (e.g. trimethylsilyl sulfide and bromide). The type of halide and precursor ratio could be adjusted to obtain either the orthorhombic BiSX or the hexagonal phase Bi$_{13}$S$_{18}$X$_2$, both of which formed nanorods due to their anisotropic crystal structures, characterized by tubular Bi-S networks. Notably, the original protocol avoided the use of amines as surfactants to prevent the reduction of Bi$^{3+}$ to metallic Bi (Figure S1-2).

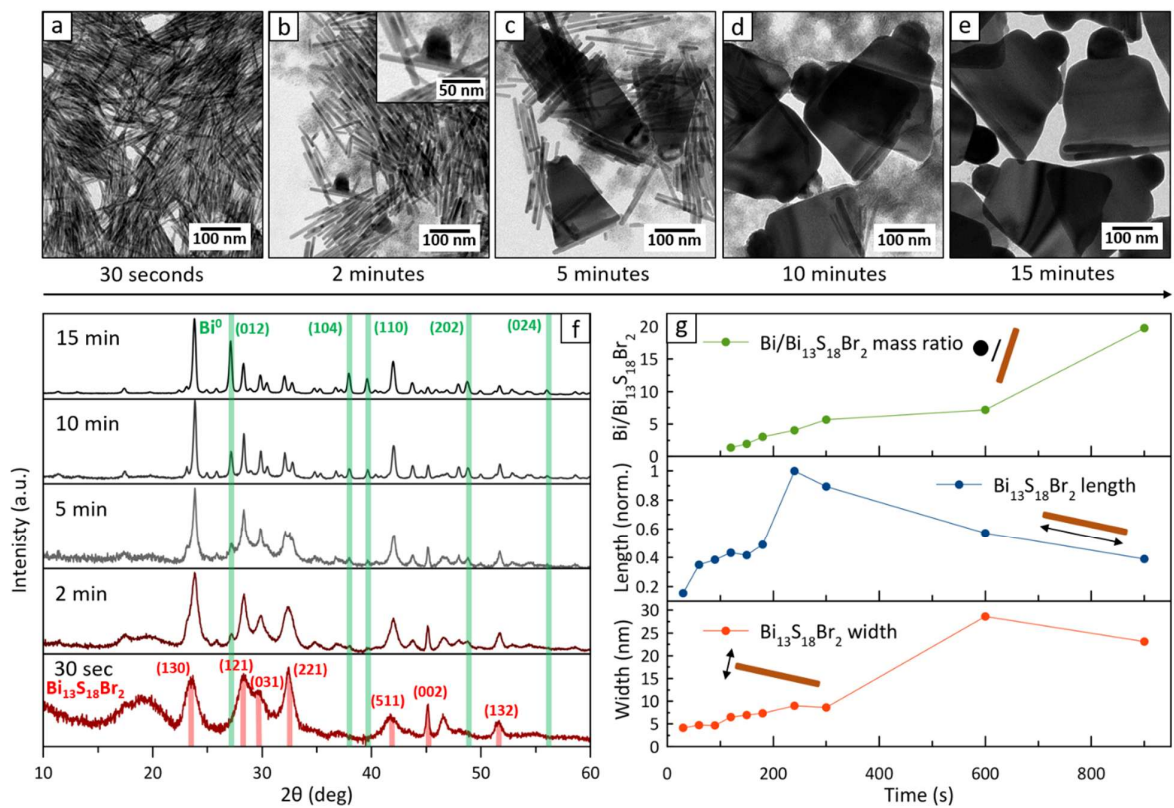

**Figure 1. Synthesis of Bi/Bi$_{13}$S$_{18}$Br$_2$ nano-bells.** a-e) TEM images of aliquots collected at 0.5, 2, 5, 10, and 15 mins from the start of the reaction. The inset in panel (b) highlights an early-stage Bi/Bi$_{13}$S$_{18}$Br$_2$ heterostructure, where the chalcohalide domain is still rod-shaped. f) XRD patterns of the aliquots, showing a progressive enlargement of the chalcohalide domains (peaks becoming sharper) and the growth of metallic-Bi domains (additional peaks appearing, marked in green). g) Compositional and morphological descriptors extracted from the Rietveld fit of XRD profiles. From top to bottom: Bi/Bi$_{13}$S$_{18}$Br$_2$ mass ratio (metallic Bi is not detectable in the first three patterns, top); average Bi$_{13}$S$_{18}$Br$_2$ crystallites length along the *c* lattice direction (rod length, middle); average Bi$_{13}$S$_{18}$Br$_2$ crystallites thickness perpendicular to the *c* lattice direction (bottom).

In the present work, we speculated that controlling this reduction process could enable the growth of functional metal-semiconductor heterostructures. Therefore, we modified the synthetic protocol for the orthorhombic BiSBr nanocrystals by injecting oleylamine into the reaction just before adding the sulfur and bromine precursors. Contrary to expectations, these initial tests yielded Bi$_{13}$S$_{18}$Br$_2$ nanorods instead of BiSBr. Among them, we observed the occasional formation of peculiar heterostructures consisting of a triangular sheet of Bi$_{13}$S$_{18}$Br$_2$ with a spherical, metallic bismuth cap attached to it (Figure S3-4), and which in projection appeared as bell-shaped.

The nucleation of $Bi_{13}S_{18}Br_2$ instead of BiSBr is consistent with a more reducing environment, as this compound contains subvalent $[Bi-Bi]^{4+}$ dimers.[34,35,36] However, the introduction of a primary amine did not lead to the formation of metallic bismuth, as initially expected. To address this, we replaced it with a tertiary amine (*N,N* - didodecylmethylamine), driven by the hypothesis that the inductive effect of the alkyl chains would make it a stronger nucleophile and a better reducing agent, capable of inducing the heterogeneous nucleation of bismuth on the surface of $Bi_{13}S_{18}Br_2$ nanorods.

Figure 1a-e illustrates the morphological evolution of the resulting particles, tracked by extracting aliquots at different time steps. In the early stages of the reaction (30 sec - 1.5 min, see Figure S5a-h) only $Bi_{13}S_{18}Br_2$ rods were visible under the transmission electron microscope (TEM). These rods were found to nucleate just after the injection of precursors, and their length steadily increased in the early phases of the reaction (Figure S6). However, after 2 min metallic-Bi domains began to form at the center of some nanorods, appearing as large and dark hemispheres (Figure 1b). As the reaction proceeded, the number of free nanorods progressively decreased while the metal-rod heterostructures grew wider, ultimately acquiring their characteristic bell shape. By 15 min, no free nanorods had remained in the sample (Figure 1d-e, see also Figure S6-8).

Rietveld fits of X-ray diffraction (XRD)[37] patterns collected from the same aliquots (Figure 1f, see also Figure S9) supported the conclusions from electron microscopy. The presence of metallic bismuth became detectable after 2 min, and its estimated relative fraction increased linearly during the reaction (Figure 1g, top). Similarly, the average lateral size of $Bi_{13}S_{18}Br_2$, extracted by matching the width of peaks with an anisotropic spherical harmonics model of the crystallite shape,[38,39,40] increased from the initial estimate of 4 nm up to ~ 25 nm. The rod length initially increased as well (up to ~4 min), indicating that the accretion of $Bi_{13}S_{18}Br_2$ nanorods was still ongoing. However, starting from ~5 min the length of such isolated nanorods dropped quickly as they started to be consumed in favor of the nano-bells, likely by an Ostwald ripening process. Note that the rod length estimates from XRD are not quantitative, as it is challenging to model strongly anisotropic and flexible particles along their longest direction[41]. However, a relative length comparison is valid, and the same morphological evolution was confirmed by TEM imaging, albeit with lower statistics (See Figure S10).

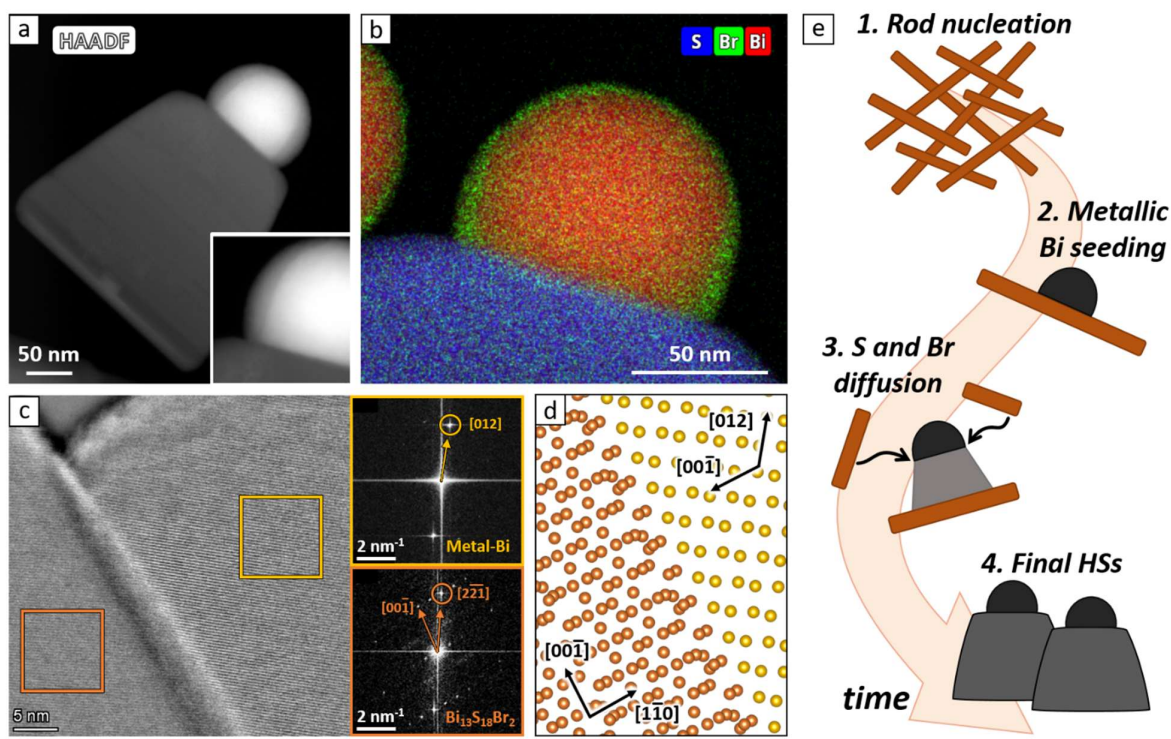

**Figure 2. Composition and structure of Bi/Bi$_{13}$S$_{18}$Br$_2$ nano-bells.** a) HAADF image of one heterostructure. b) STEM-EDX compositional map of the Bi/chalcohalide contact region, showing the presence of a Br-rich shell surrounding the metal hemisphere. c) HAADF image of the Bi/Bi$_{13}$S$_{18}$Br$_2$ interface region, with lattice fringes enhanced by a high-pass filter. Insets: Fourier transforms of the yellow and orange areas respectively. The circled spots identify the most prominent lattice planes visible along this orientation. d) Proposed Bi/Bi$_{13}$S$_{18}$Br$_2$ epitaxial model of the interface, with the two domains oriented in accordance with panel (c). Only bismuth atoms are shown for clarity, and the relative position of the two domains is estimated (i.e., no model optimization performed). e) Proposed formation mechanism for the Bi/Bi$_{13}$S$_{18}$Br$_2$ nano-bells.

Scanning-transmission electron microscopy images (STEM, Figure 2a) and energy-dispersive X-ray spectroscopy (EDX, Figure 2b) maps of the heterostructures evidenced that Bi, S and Br atoms are uniformly distributed in the chalcohalides domain (Figure S11) and revealed the presence of a ~1.3 nm Br-rich layer on the surface of the metallic-Bi hemisphere. This layer grows thicker (~ 7-8 nm) at the neck region connecting the two domains and is likely polycrystalline (Figure S12). These observations are congruent with the elemental composition observed in each domain by EDX analysis (Figure S13). Further inspection of the interface by high-resolution (HR)-STEM and 4D-STEM suggested an epitaxial relation between the two domains, as indicated by the matching periodicity of the lattice fringes at the interface and the presence of nearly aligned atomic planes (Figure 2c). A plausible epitaxial relation is (001)//(1-10) – Bi/Bi$_{13}$S$_{18}$Br$_2$, which was identified by feeding the probable lattice orientations inferred by 4D-STEM into the lattice matching module of the Ogre library for the prediction of epitaxial interfaces,[42] developed by the Marom group in collaboration with some of us (Figure 2d, see Figure S14 for further discussion).

When combined, these observations allow outlining a possible growth mechanism for the Bi/Bi$_{13}$S$_{18}$Br$_2$ nano-bells (Figure 2e). Initially, the injection of sulfur and bromine precursors triggers the nucleation of Bi$_{13}$S$_{18}$Br$_2$ nanorods, driven by the reactivity of the precursors and the ionic nature of the compound. Shortly thereafter, the amine starts reducing Bi$^{3+}$ to metallic bismuth, a process facilitated by the nanorods serving as heterogeneous nucleation templates. This reduction proceeds gradually due to the tertiary amine's weak reducing ability and its consumption during the reaction.

Once the Bi/Bi$_{13}$S$_{18}$Br$_2$ interface forms, it can act as a sink for excess anions. Bromine is already found at the surface of the metal hemisphere, and sulfur is adsorbed to produce more Bi$_{13}$S$_{18}$Br$_2$ at the interface. This enlarges the chalcohalide sheet and gradually displaces the original nanorod (now forming the bottom edge of the bell) away from the bismuth hemisphere. Similar growth mechanisms are well documented in nanofabrication via molecular beam epitaxy, such as the use of Au nanoparticles to drive the grow of GaAs nanopillars,[43,44,45] but remain rare in colloidal heterostructures[46]. The process continues until all the precursors in solution are depleted (~5 min). At this point, Ostwald ripening drives a further transfer of material from the nanorods to the heterostructures. As a result, the remaining metal-free nanorods gradually shrink and eventually disappear, leaving behind only Bi/Bi$_{13}$S$_{18}$Br$_2$ nano-bells at the end of the reaction.

To further validate this mechanism we combined two reaction batches, one where the bells were allowed to fully develop and another that was quenched at the nanorods stage. The mixture was heated back to 180 °C to resume the reaction, which proceeded for an additional 15 minutes. As expected, the nanorods from the second batch did not develop into new, small heterostructures; instead, they were consumed to further enlarge the existing ones (see Figure S15).

We note that the epitaxial relation proposed in Figure 2d should cause an expansion of the chalcohalide in the plane of the interface (see Figure S14). A 4D-STEM analysis, performed by capturing local electron diffraction patterns while scanning across one heterostructure, confirmed the presence of a strain field propagating from the Bi/Bi$_{13}$S$_{18}$Br$_2$ interface. Figure 3b-e shows the linear strain components ($\epsilon_{xx}$ and $\epsilon_{yy}$), along with the shear strain and local lattice rotation ($\epsilon_{xy}$ and $\theta$), reconstructed with py4DSTEM.[47] The analysis confirms an $\epsilon_{yy}$ expansion of the $Bi_{13}S_{18}Br_2$ lattice in the interface plane and a corresponding orthogonal $\epsilon_{xx}$ compression to keep the unit cell volume unchanged, in agreement with our predictions. Interestingly, while performing a HAADF tilt series we also found that the darker domain often visible in TEM at the bottom of the bells is not a local thickening, but rather a curl that forms as the chalcohalide sheet rolls up at its bottom edge (Figures 3f). We speculate that this behavior might result from the need to accommodate some of the interface strain, and is likely facilitated by the hexagonal crystal structure of $Bi_{13}S_{18}Br_2$, which allows the domain to expose the same surface at the interface after a 60° rotation.

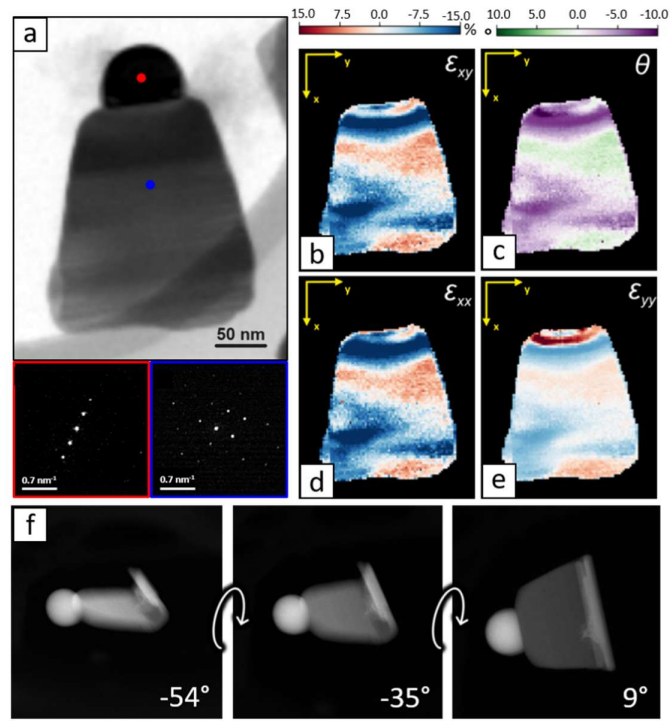

**Figure 3. 4D-STEM and tomography analysis.** a) Virtual bright field image reconstructed from the diffraction patterns of Bi and $Bi_{13}S_{18}Br_2$ collected at each pixel. b-c) Map of the shear strain and rotation of the lattice. d-e) $Bi_{13}S_{18}Br_2$ strain field components along the x- and y-directions. The x and y axes in panels (b-e) indicate the images reference system, and are not related to crystallographic directions. f) Frames from a HAADF tilt series, showing curling of the lower section of a bell.

Owing to the 1.65 eV band gap[48] of $Bi_{13}S_{18}Br_2$, its high absorption coefficient,[49] and its intimate connection with the metallic-Bi domain, these heterostructures are interesting candidates for converting visible light into highly accessible photocarriers. This motivated us to screen the photocatalytic activity of $Bi/Bi_{13}S_{18}Br_2$ nano-bells. As a model reaction, we selected the photodegradation of Rhodamine B (RhB), for which other Bi-based semiconductors are known to be active.[50,51] The reaction was conducted in isopropanol under 420 nm irradiation (see Methods), resulting in the complete degradation of RhB after 1 h. Control tests performed in the absence of heterostructures confirmed that the intrinsic photodegradation of the dye was negligible under the conditions adopted (4.5 %, Figure S16). Likewise, tests performed under dark conditions, *i.e.*, letting the RhB surface adsorption equilibrium establish, and then removing the particles resulted in a negligible decrease of the absorption signal (< 1 % RhB adsorption on the HS, Figure S17), confirming that the observed dye degradation was indeed the result of a photocatalytic process mediated by the $Bi/Bi_{13}S_{18}Br_2$ heterostructures.

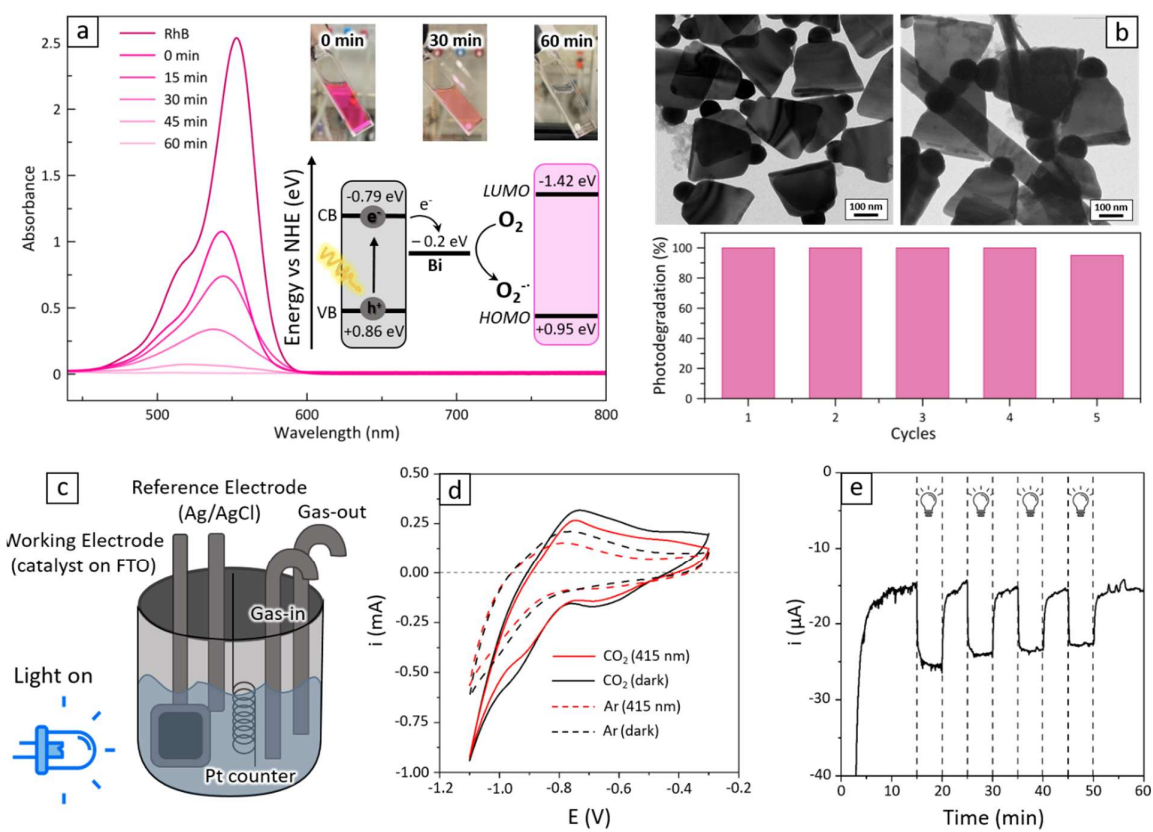

**Figure 4. Photo- and photoelectro-catalytic activity of $Bi/Bi_{13}S_{18}Br_2$ heterostructures.** a) Absorption spectra and pictures (insets) of the RhB solution taken at various stages of the photodegradation reaction (top) and scheme of the proposed photodegradation mechanism (down). b) TEM images of Bi HSs before (left) and after (right) 5 cycles of RhB photodegradation (top); photodegradation activity for the five cycles (down). c) Scheme of the cell used for the photoelectrochemical testing of heterostructure-based electrodes. (d) Cyclic voltammetry traces collected on the heterostructures

($v$ = 100 mV s$^{-1}$). The dashed curves have been collected under Ar bubbling (blank tests, only hydrogen evolution reaction is possible), while the full traces are related to CO$_2$ reduction reaction tests. Black: dark conditions. Red: 415 nm LED illumination. (e) Chronoamperometric scan, performed at -1 V vs the reference electrode under intermittent illumination.

The photocatalytic mechanism was further investigated by repeating the reaction in the presence of benzoquinone, which acts as a scavenger for the superoxide ion (O$_2^{\cdot-}$).[52] Under these conditions, the RhB photodegradation was reduced to only 5.4 % (Figure S18), highlighting a key role of O$_2^{\cdot-}$ in the process. Indeed, a second test conducted in oxygen-free conditions to prevent the superoxide anion formation (N$_2$ atmosphere, see Figure S19) resulted in almost no photodegradation of RhB. Note that we excluded the ·OH radical as a possible reactive species, because isopropanol is an efficient ·OH scavenger[53,54]. Figure S20 summarizes the result of control experiments performed in the presence of N$_2$ and benzoquinone, in the absence of photocatalyst, and in the dark.

Based on this evidence, we propose the reaction mechanism illustrated in Figure 4a (inset). First, in line with the Bi/Bi$_{13}$S$_{18}$Br$_2$ band alignment, the photogenerated electron moves to the metallic-Bi domain, which acts as a cocatalyst. There, the electron is unlikely to recombine with the hole, which remains localized on the chalcohalide domain and becomes accessible to reduce one molecule of O$_2$ to the O$_2^-$ superoxide anion O$_2^{\cdot-}$, which then oxidizes the substrate[55,56]. The fate of the hole is uncertain: it may be consumed by reducing the isopropanol or may further activate the superoxide by transforming it into singlet-oxygen ($^1$O$_2$), which is itself a highly reactive oxidizer. The effectiveness of benzoquinone as O$_2^{\cdot-}$ scavenger reveals that this is the main reactive oxygen species (ROS) responsible for photodegradation of RhB. The photodegradation mechanism operated by this reactive oxygen species (ROS) could involve N-de-ethylation, chromophore cleavage and eventually mineralization steps. Based on the hispochromic shift observed in the absorption spectra during the irradiation time ($\Delta\lambda$ = 16 nm at 45 min), we can hypothesize the N-de-ethylation of the dye with the formation of a series of N-de-ethylated intermediates as a possible mechanism. This could be followed by the cleavage of the chromophore structure, suggesting a stepwise breakdown of the molecular structure of the dye, in agreement with previous reports.[57] On the other hand, the use of IPA as solvent ruled out the participation of OH· radicals and holes (h$^+$) into the mechanism, as it is a well-known scavenger of both species[58].

Further tests performed with methylene blue showed similar results to RhB, with 93% degradation after 2 hours (Figure S21-23), while no degradation was observed for methyl orange (Figure S24).

This difference is attributed to a poor affinity of the anionic methyl-orange for the nano-bell surface, compared to cationic dyes like RhB and methylene blue, which is consistent with the measured negative Zeta potential value (-58.3 mV, see Table S1).

Notably, the heterostructures maintained almost full photocatalytic activity (ca. 95% degradation of the selected dyes) after 5 cycles (Figure 4b), with virtually no catalyst degradation and only minor signs of aggregation, despite the prolonged exposure to a polar environment. Although the colloidal stability of the nanocrystals is limited due to their large size, a post-reaction TEM analysis showed no noticeable alteration of the material, further attesting its robustness (Figure 4b). This stability underscores the potential of these heterostructures for practical applications where durability and consistent performance are required.

Motivated by this remarkable durability, we attempted a preliminary test for the activity of our Bi/$Bi_{13}S_{18}Br_2$ heterostructures in the $CO_2$ reduction reaction ($CO_2$RR). Indeed, metallic-Bi is known to be a good catalyst for $CO_2$RR, and the heterostructures' resilience to a polar solvent like IPA suggested they could withstand water.

The $CO_2$RR tests were conducted using our previously reported electrochemical $CO_2$RR setup.[59] Briefly, photoelectrodes were crafted by spin coating the heterostructures on fluorine-doped tin-oxide (FTO) glass, and were installed in a three-electrode, single-compartment quartz cell (see Figure 4c and methods for setup, electrochemical testing, and product quantification). Preliminary Open Circuit Voltage (OCP, Figure S25) measurements performed under intermittent 415 nm illumination revealed a photo-response in the form of a reversible shift towards more cathodic values upon illumination. Conversely, Cyclic Voltammetry (CV)[60] displayed limited sensitivity to light in terms of current/voltage relationship (black versus red traces in Figure 4d). For reference, a comparable signal was measured from the bare FTO control electrode (Figure S26).

Nevertheless, clear differences in the CV traces emerged when $CO_2$ was introduced in the setup, indicating catalytic activity. As shown by the dashed traces in Figure 4d (collected under Ar flow), a smooth trace with a clear onset potential ($E_{Onset}$) around -0.8 V vs RE can be observed and is attributed to the Hydrogen Evolution Reaction (HER). In contrast, when $CO_2$ is present (full lines in Figure 4d), the voltammogram displays additional features (peaks/waves at ca. -0.6 and -1 V vs RE) that are consistent with the $CO_2$ pre-adsorption and reduction onset. This hypothesis is supported by the increased current density recorded at large cathodic potentials, which indicates a $CO_2$RR contribution to the HER current observed in Ar-fed tests.

In an attempt to identify parasitic HER and CO$_2$RR product distribution and Faradaic efficiency, we also performed a 1 hour-long ChronoAmperometric (CA) scan at -1 V vs RE under CO$_2$RR conditions. The intermittent illumination (5 minutes-long on/off cycles, Figure 4e) clearly enhanced the reaction rate (*i.e.*, current density), further proving the CO$_2$RR photoelectroactivity of heterostructures, as the same photoresponse was not observed during control tests (Figure S27). However, the low current densities translated into limited production rates, making it challenging to detect and quantify both HER and CO$_2$RR products (see Figure S28 and related discussion).

In conclusion, we have reported a synthetic protocol to obtain Bi/Bi$_{13}$S$_{18}$Br$_2$ metal/semiconductor heterostructure through the controlled introduction of a reducing agent. These heterostructures adopt a unique bell-like morphology, resulting from a complex, multi-stage nucleation process: the formation of initial Bi$_{13}$S$_{18}$Br$_2$ nanorods, followed by the growth of metallic-Bi on their surface, and finally by an interfacial accretion process driven by Ostwald ripening. These heterostructures demonstrate remarkable stability under operando conditions, even when exposed to polar solvents like isopropanol and water. They also show high photocatalytic activity, good recyclability under visible-light irradiation for cationic dyes photodegradation, and preliminary CO$_2$ reduction capabilities. Their ease of synthesis, use of non-toxic and earth-abundant elements, and promising performances highlight their potential for future applications in and beyond photocatalysis.

**Chemicals**

All chemicals were of the highest purity available unless otherwise noted and were used as received. Bismuth (III) acetate (Bi(Ac)3; 99.99 %), oleic acid (technical grade, 90 %), 1-octadecene (technical grade, 90 %), bis(trimethylsilyl)sulfide ((Me3Si)2S; synthesis grade) and trimethylsilyl bromide ((Me3Si)Br; 97 %) were purchased from Sigma-Aldrich. *N,N* - Didodecylmethylamine (85%) was purchased from TCI. Hexane and toluene were purchased from Sigma-Aldrich. All solvents were purchased anhydrous and were used as received.

**Synthesis of Bi/Bi$_{13}$S$_{18}$Br$_2$ nano-bells**

The synthesis is a modification of the already reported method by Giansante et al.[61] In short: all colloidal nanocrystals (NCs) were synthesized in three-necked flasks attached to a standard Schlenk line, ensuring oxygen- and moisture-free conditions. In a typical procedure to obtain Bi / Bi$_{13}$S$_{18}$Br$_2$ NCs, 0.3 mmol (120 mg) of Bi(Ac)$_3$ and 6.3 mmol (2 mL) of oleic acid were combined in 3 mL of 1-octadecene. The mixture was stirred vigorously and deoxygenated through several cycles of vacuum and nitrogen purging at around 80 °C. Subsequently, the mixture was heated to 110 °C to fully dissolve Bi(Ac)$_3$, which caused the solution to turn colorless and optically clear, indicating the complete formation of bismuth(III)-oleate complex(es). The solution was maintained under vacuum at 110 °C for 30 minutes to eliminate any acetic acid formed during the complexation process. Afterward, the solution was reheated under nitrogen flow, with the temperature stabilized at 180°C. At this stage, 300 µL of didodecylmethylamine was injected, followed immediately by the co-injection of a half equivalent of (Me$_3$Si)$_2$S (0.15 mmol; 32 µL) and one equivalent of (Me$_3$Si)Br (0.3 mmol; 39.5 µL) dissolved in 2 mL of octadecene. The reaction was allowed to proceed for 15 minutes before the heating source was removed. The colloidal dispersion was quickly cooled to room temperature by placing the reaction flask in an ice bath. After the reaction, the mixture was centrifuged at 6000 rpm for 5 minutes. The clear supernatant was discarded, and the resulting black precipitate was collected in 1 mL of toluene or hexane. No additional purification steps were carried out on the NC heterostructures.

**Electron microscopy**

Bright Field Transmission Electron Microscopy (BF-TEM) measurements of the NCs were performed using either a JEOL JEM-1011 with a W thermionic source at an acceleration voltage of 100 kV or a JEOL JEM-1400Plus TEM, with LaB$_6$ thermionic source and maximum acceleration voltage 120 kV. The highly diluted NC solution was first put in ultrasound for 1 minute and then drop-cast onto copper grids (200 mesh) with carbon film and the solvent was then allowed to evaporate in a vapor-controlled environment. The longitudinal and lateral dimensions were assessed through statistical analysis of TEM images of several hundred NCs using the ImageJ software. Selected area electron diffraction measurements were performed on the same microscope and evaluated using the CrysTBox software package.

High-resolution scanning transmission electron microscopy (HRSTEM) images were acquired on a probe-corrected Thermo Fisher Spectra 300 STEM operated at 300 kV. Images were acquired on a high-angle annular dark-field (HAADF) detector with a current of ~100 pA. Compositional maps were acquired and analyzed using Velox, with a probe current of ~150 pA and rapid rastered scanning, on a Dual-X detector setup. 4DSTEM data were collected on a Gatan Continuum using STEM-X, with a current of ~5 pA and a dwell time of 1 ms. The analysis of the 4DSTEM datasets was carried out using the open-source Python code py4DSTEM[62].

**X-ray Powder Diffraction**

Characterization by X-ray Powder Diffraction (XRD) was performed by employing a PANalytical Empyrean X-ray diffractometer using a 1.8 kV Cu Kα ceramic X-ray tube operating at 45 kV and 40 mA and detected by a PIXcel3D 2 × 2 area detector. Samples were prepared by dropcasting highly concentrated solutions on zero-diffraction silicon substrates. All diffraction patterns were acquired at room temperature under ambient conditions. Data analysis was performed using the HighScore 4.9 software from PANalytical. The Rietveld fits of XRD patterns were performed with the FullProf suite[63], and were specifically oriented to the extraction of morphological information and mass ratios. To this end, additional effort was devoted to the appropriate modelling of the peak intensities and profiles. The instrumental broadening was quantified by measuring and fitting a $LaB_6$ diffraction pattern (not shown) to construct the instrumental response function. Then, the pattern was fitted by refining for both phases (metallic Bismuth and $Bi_{13}S_{18}Br_2$) the following parameters: scale factor, lattice parameters ($a,b,c$, whereas angles were fixed by symmetry), polynomial background, instrumental zero, and spherical harmonics size parameters (1 parameter for metallic Bismuth = spherical crystallite, up to 3 parameters for $Bi_{13}S_{18}Br_2$ = Y00, Y20, Y40). March-Dollase modelling for the preferred orientation of $Bi_{13}S_{18}Br_2$ was included to account for the possible orientation of the large and flat nano-bells upon deposition on the substrate, to correctly account for the observed intensity of reflections. Instead, the position of atoms was not refined to avoid the introduction of unjustified distortion to the crystal structure, which would likely mask the actual underlying microstructural contributions to the pattern. In those cases where the Bi-domain was close to its detection limit, the domain size was fixed to the last value that could be determined reliably to avoid overfitting.

**Photocatalysis tests**

For the photodegradation of the organic dyes, a 1 mM solution of MB or RhB was prepared in 20 mL of IPA, in 20 mL of methanol in the case of MO. Then, 2 mg of $Bi/Bi_{13}S_{18}Br_2$ NCs were mixed with 73 µL of the organic solution until a final volume of 2.5 mL of IPA in quartz cuvettes. Subsequently, the cuvettes were introduced in a photoreactor (λex=420 nm) until the complete photodegradation of the organic pollutants, measuring the UV-Vis spectra for following the process at different photoreaction times. The solution containing the photocatalyst and the organic molecule was centrifuged (7000 rpm for 5 minutes), and the spectra were collected from the supernatant. UV–visible absorption spectra were recorded on a spectrophotometer UV/VIS/NIR Lambda 1050, equipped with software PerkinElmer UV Winlab. Quartz cuvettes of 1 cm × 1 cm path length were used to acquire all the data. As photoreactor was used the PhotoreactorM2 from Penn PhD, adjusting the stirring at 400 rpm and the fan at 2800 rpm. For the study of the reaction mechanism, a 1 mM solution of benzoquinone (scavenger) was prepared in 1 mL of IPA. Then, 2 mg of Bi/Bi$_{13}$S$_{18}$Br$_2$ NCs were mixed with 73 µL of RhB solution and 100 µL of BZQ solution until a final volume of 2.5 mL of IPA in quartz cuvette. Before the irradiation, the mixture was stirred for 15 minutes in the dark to establish the adsorption-desorption equilibrium of RhB on the photocatalyst surface. The zeta (ζ) potential was determined by employing a Zetasizer Ultra (Malvern, UK).

**Photoelectrochemical tests**

<u>Photoelectrodes preparation</u>. 1x1 cm electrodes where crafted starting from the toluene suspensions of Bi HSs and Bi rods obtained from the synthesis. They were obtained by spin coating (Laurell WS-650MZ-23NPPB spin coater, equipped with a GAST 0523-101Q-G588NDX Vacuum Pump) on FTO glass. Optimal film homogeneity was obtained operating at 5000 RPM for 1 minute, with either 3x 50 µL (Bi HSs).

<u>Photoelectrochemical setup</u>. The photoelectrochemical characterization of the sample has been carried out in a single compartment quartz cell, purchased from PineResearch ("Low Volume Photoelectrochemical Three Electrode Quartz Cell"). Tests were performed under a typical three-electrode configuration, using the above-described photoelectrodes as working electrode (WE), a Pt coil as counter electrode (CE) and an Ag wire (sealed in a fritted glass tube, filled with the working electrolyte) as reference electrode (RE). Gas inlet and outlets were also implemented. The whole cell was gas-tight, with the gas outlet connected to an inline gas chromatograph (GC). The testing electrolyte was a 0.1 M LiClO$_4$ solution in ACN:H$_2$O (99:1 v/v). A THORLABS M415L4 - 415 nm, 1310 mW (Min) Mounted LED, 1500 mA was employed as illumination source and an Ivium Compact Stat.h as the potentiostat. The tests were performed feeding the cell by bubbling Ar (delivered by a Bronkhorst F-201CV-100-RGD-22-V mass flow controller) or CO$_2$ (delivered by a Bronkhorst F-201CV-100-RGD-22-V mass flow controller), depending on the nature of the test, with a flow rate set at 5.5 sccm. The gas outlet was connected in-line to a gas dryer (SRI Gas Stream Dryer, SRI part# 8670-5850), to a universal Mass Flow Meter (EL-Flow Prestige Bronkhorst, model FG-111B-AGD-22-V-DA-000) and then to the GC for the quantification of gaseous products.

Electrochemical tests. Three main electrochemical techniques have been used in the characterization of the CO$_2$RR photoelectroactivity of the samples under study. Open Circuit Potential (OCP) was measured under illumination and/or dark conditions, as to assess the interaction of the thin films with the incident radiation. Cyclic Voltammetry (CV) has been also operated under both illumination and dark conditions, in the presence and absence of CO$_2$, depending on the nature of the test, aiming at assessing the CO$_2$RR features of the materials. Finally, ChronoAmperometric scans (CA, 1 hour-long) have been recorded, again under both illumination and dark conditions, as to determine the CO$_2$RR product distribution yielded by the materials at definite potentials. Unless otherwise stated, all the potentials reported in this work are measured versus the previously described RE. Blank tests were performed on the electrode support (*i.e.*, bare FTO, 1x1 cm).

Products quantification. Gaseous products were detected by an inline-connected SRI 8610C gas chromatograph (Multiple Gas Analyzer #5) equipped with a thermal conductivity detector (TCD) and flame ionization detector (FID) coupled with a methanizer. Argon has been used as gas carrier. Detectors response have been calibrated prior to experiments using a customized gas mixture containing the most typical CO$_2$RR gas phase products. Details on calibration and quantification methods can be found in our previous works.[64] Liquid products were detected ex-situ, sampling the post-reaction electrolytes, by HPLC. All HPLC separations have been carried out on an Agilent Infinity 1260, equipped with a quaternary pump, standard autosampler, thermostatted column compartment, Diode Array Detector (DAD/UV-Vis) operating in the range between 190 to 950 nm and a Refractive Index Detector (RID). In particular, possible CO$_2$RR liquid products (*e.g.*, formic acid/formate ions) have been separated on an Agilent Hi-Plex H (300 × 7.7 mm) operated at 50°C in 5 mM M H$_2$SO$_{4(aq)}$, isocratic mode, 0.6 mL min$^{-1}$, 25 µL injection volume. Chromatographic runs lasted for 30 minutes, allowing for the elution of all common CO$_2$RR products. Products were detected by either DAD/UV-Vis (210 and 280 nm) and/or RID (T = 40°C) depending on the nature of the analyte. A complete overview on the separation of liquid CO$_2$RR products according to this methodology can be found in literature.[65]

**Supporting Information**

Supporting figures including TEM images and XRD pattern of synthesis with and without oleylamine; TEM images, XRD Rietveld fits and size evolution of aliquots at different reaction time; HAADF image and STEM-EDX compositional map of heterostructures; expanded description of epitaxial interface identified with Ogre; UV spectra and photographs of photocatalysis experiments; table of Z

potential; open circuit voltage measures, cyclic voltammetry, chronoamperometry, GC and HPLC chromatographs of blank and heterostructures samples.

## Acknowledgments

P.R., S.T. and L.M. acknowledge funding from the Project IEMAP (Italian Energy Materials Acceleration Platform) within the Italian Research Program ENEA-MASE (Ministero dell'Ambiente e della Sicurezza Energetica) 2021-2024 "Mission Innovation" (agreement 21A033302 GU n. 133/5-6-2021). L.M. acknowledges funding from European Research Council through the ERC Advanced Grant NEHA (grant agreement n. 101095974). J.P-P and R.G acknowledge funding from Generalitat Valenciana (CIPROM/2022/57) and J.P-P acknowledges funding from Generalitat Valenciana (IDIFEDER/2018/064, IDIFEDER/2021/064). This study forms part of the Advanced Materials programme (MFA/2022/051) and was supported by MICIN with funding from European Union NextGenerationEU (PRTR-C17.I1) and by Generalitat Valenciana.

# Synthesis, Growth Mechanism, and Photocatalytic Properties of Metallic-Bi/Bi$_{13}$S$_{18}$Br$_2$ Nano-Bell Heterostructures

Anna Cabona[1,2], Stefano Toso[1,3], Andrea Griesi[4], Martina Rizzo[1,2], Michele Ferri[1], Pascal Rusch[1], Giorgio Divitini[4], Julia Pérez-Prieto*[5], Raquel E. Galian*[5], Ilka Kriegel*[2], Liberato Manna*[1]

[1] Nanochemistry Department, Italian Institute of Technology, Via Morego 30, 16163 Genova, Italy
[2] Department of Applied Science and Technology, Politecnico di Torino, Corso Duca degli Abruzzi 34, 10129 Turin, Italy
[3] Lund University, Division of Chemical Physics, Naturvetarvägen 14, 221 00 Lund, Sweden
[4] Electron Spectroscopy and Nanoscopy, Italian Institute of Technology, Via Morego 30, 16163 Genoa, Italy
[5] Institute of Molecular Science, University of Valencia, c/Catedrático José Beltrán Martínez 2, 46980 Paterna, Valencia, Spain

## SUPPORTING INFORMATION

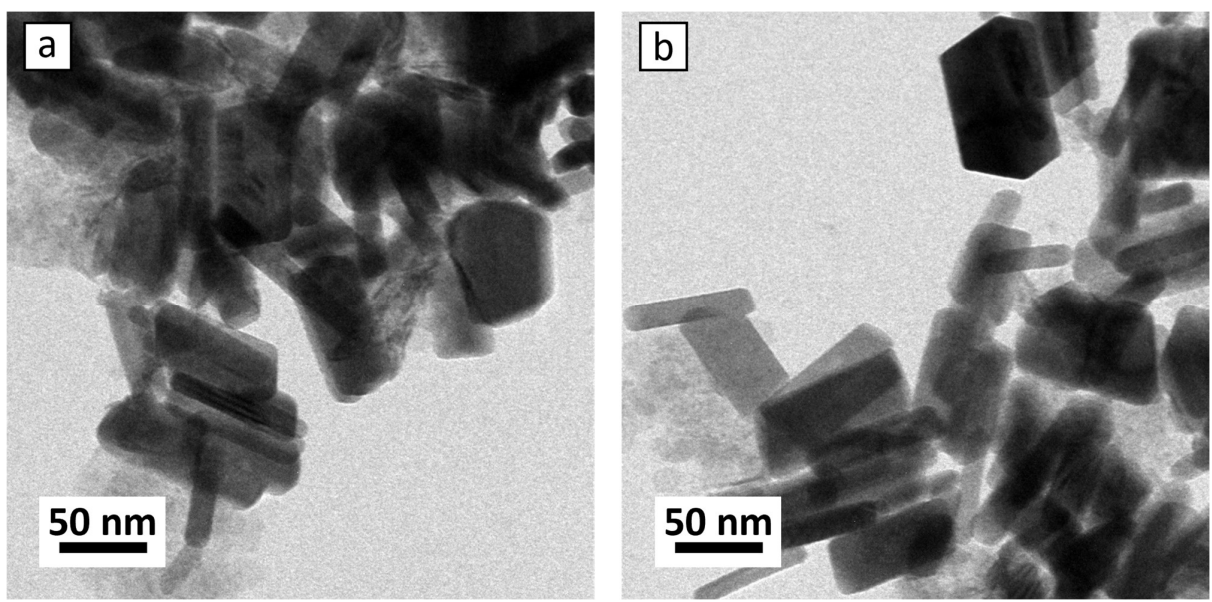

**Figure S1. Control synthesis in the absence of amines.** a-b) TEM images of colloidal BiSBr NCs synthetized in the absence of amines, following the procedure reported by Quarta et al.[65]

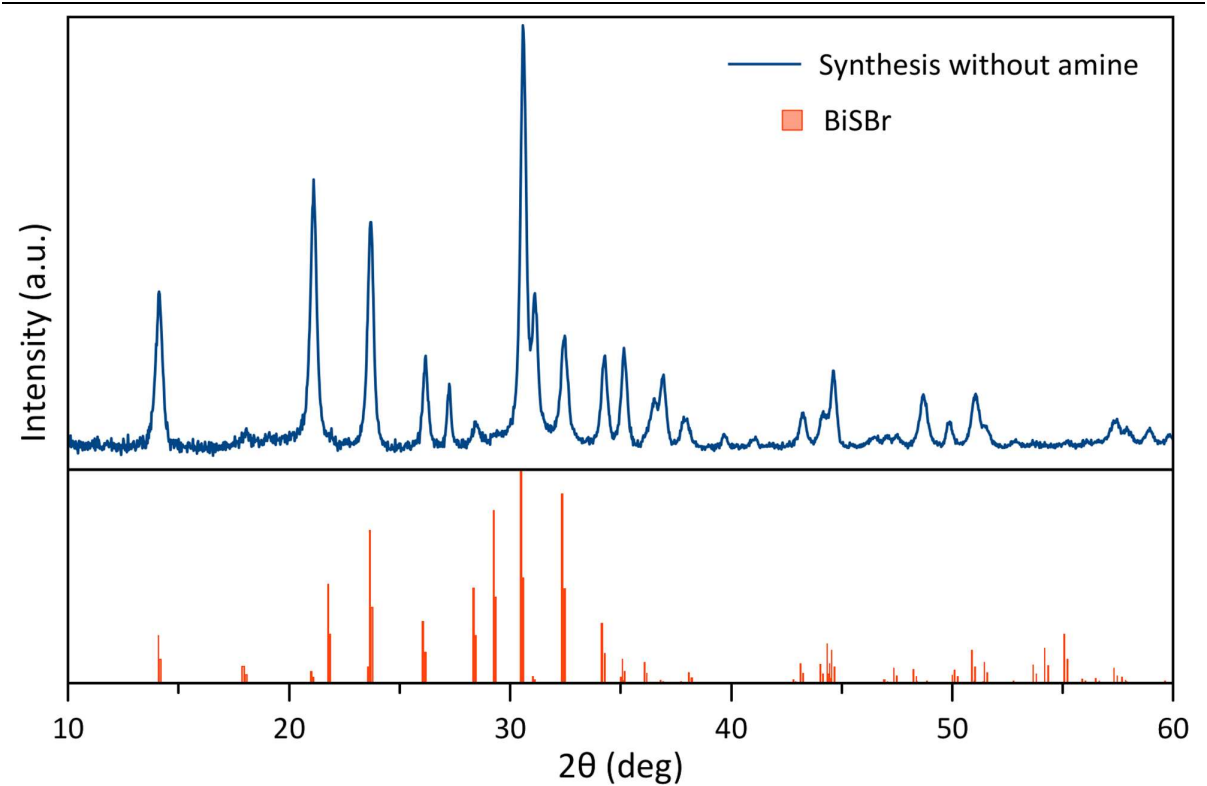

**Figure S2. Control synthesis in the absence of amines.** XRD pattern of colloidal BiSBr NCs synthetized in the absence of amine, confirming the phase attribution to BiSBr.

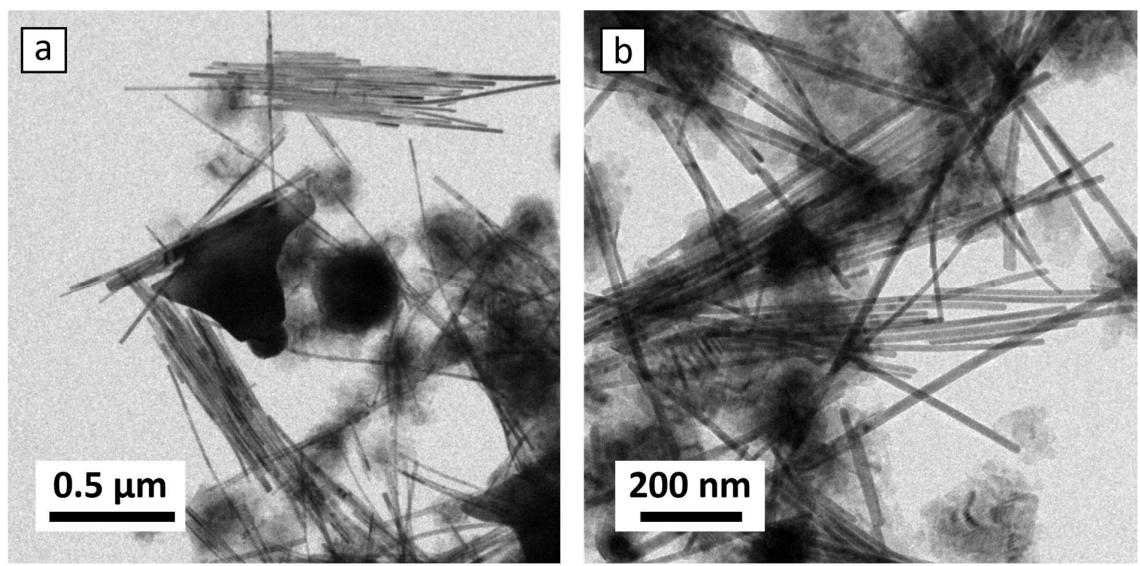

**Figure S3. Synthesis performed with added oleylamine.** TEM images of colloidal $Bi_{13}S_{18}Br_2$ obtained by adding 220 µL of oleylamine on top of the reagents used for Figures S1-2 while adopting identical reaction conditions.

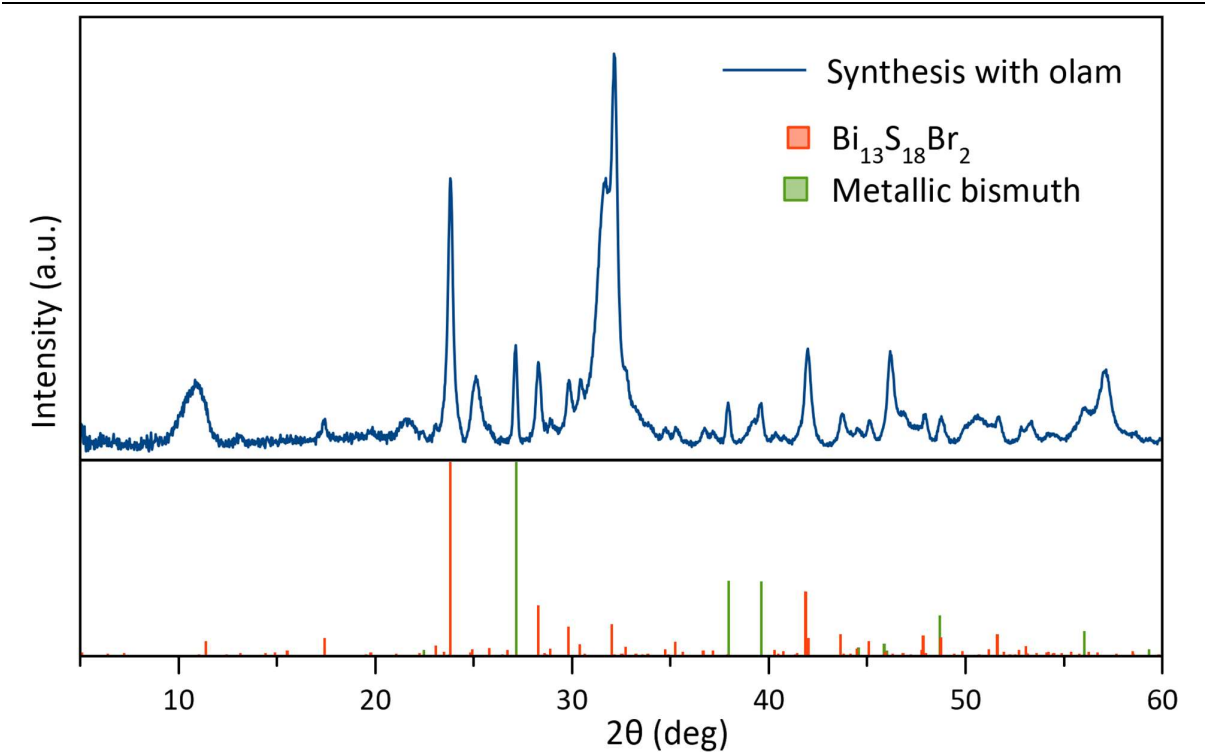

**Figure S4. Synthesis performed with added oleylamine.** XRD pattern of the colloidal Bi$_{13}$S$_{18}$Br$_2$ nanorods obtained by adding oleylamine (see Figure S3).

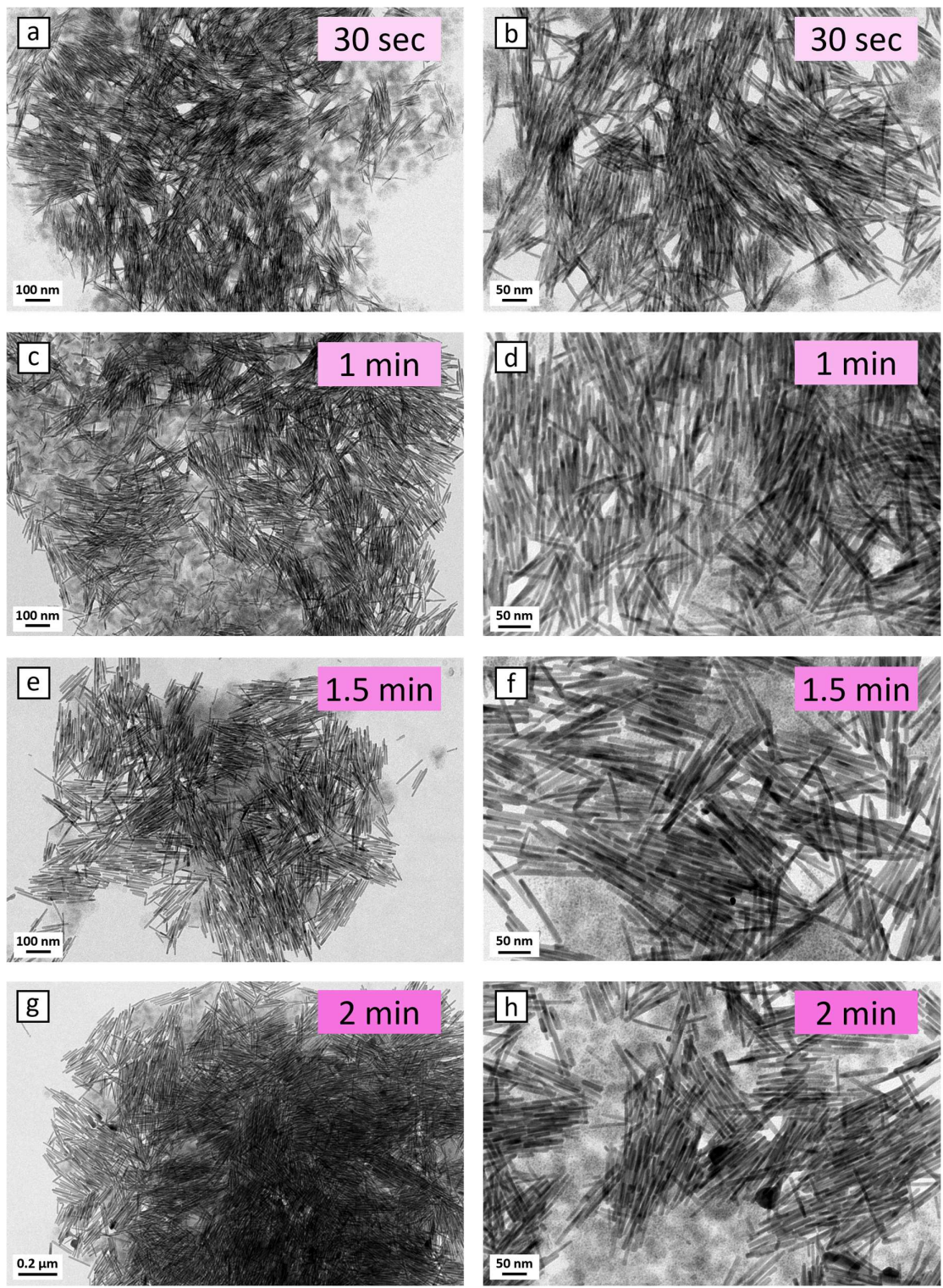

**Figure S5. Morphology evolution.** TEM images of Bi/Bi$_{13}$S$_{18}$Br$_2$ solution aliquots taken at 30 s (a-b), 60 s (c-d), 1.5 min (e-f), and 2 min (g-h) after the start of the reaction.

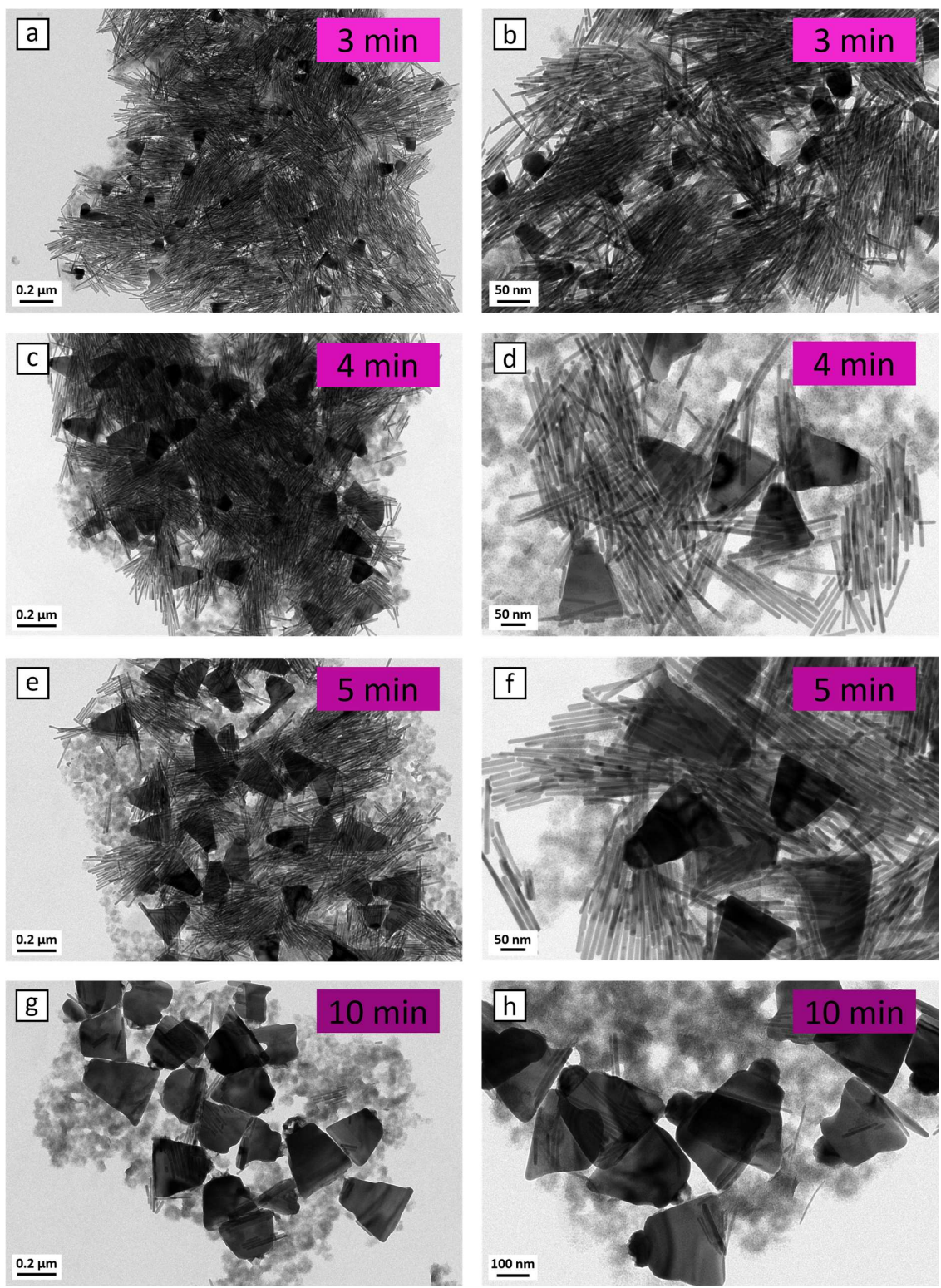

**Figure S6. Morphology evolution (continued).** TEM images of Bi/Bi$_{13}$S$_{18}$Br$_2$ solution aliquots taken at 3 min (a-b), 4 min (c-d), 5 min (e-f), and 10 min (g-h) after the start of the reaction.

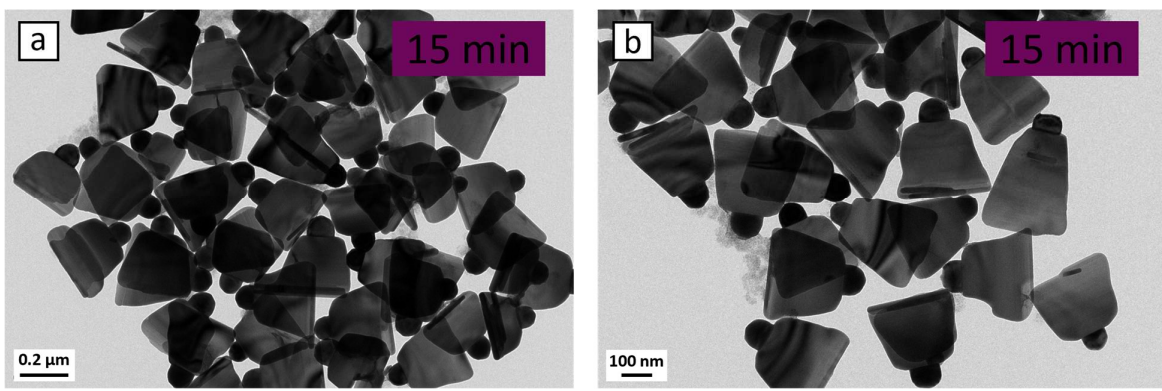

**Figure S7. Morphology evolution (continued).** TEM images of Bi/Bi$_{13}$S$_{18}$Br$_2$ solution aliquots taken at 15 min (a-b) after the start of the reaction.

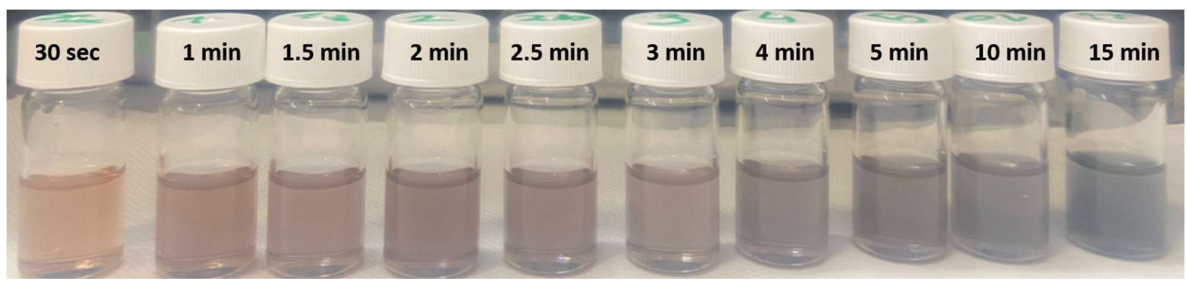

**Figure S8. Aliquot color evolution.** Color of the reaction medium aliquots as the reaction progresses. The solutions were diluted 1:75 to enhance the visibility of these color changes and to prepare TEM grids with well-separated particles.

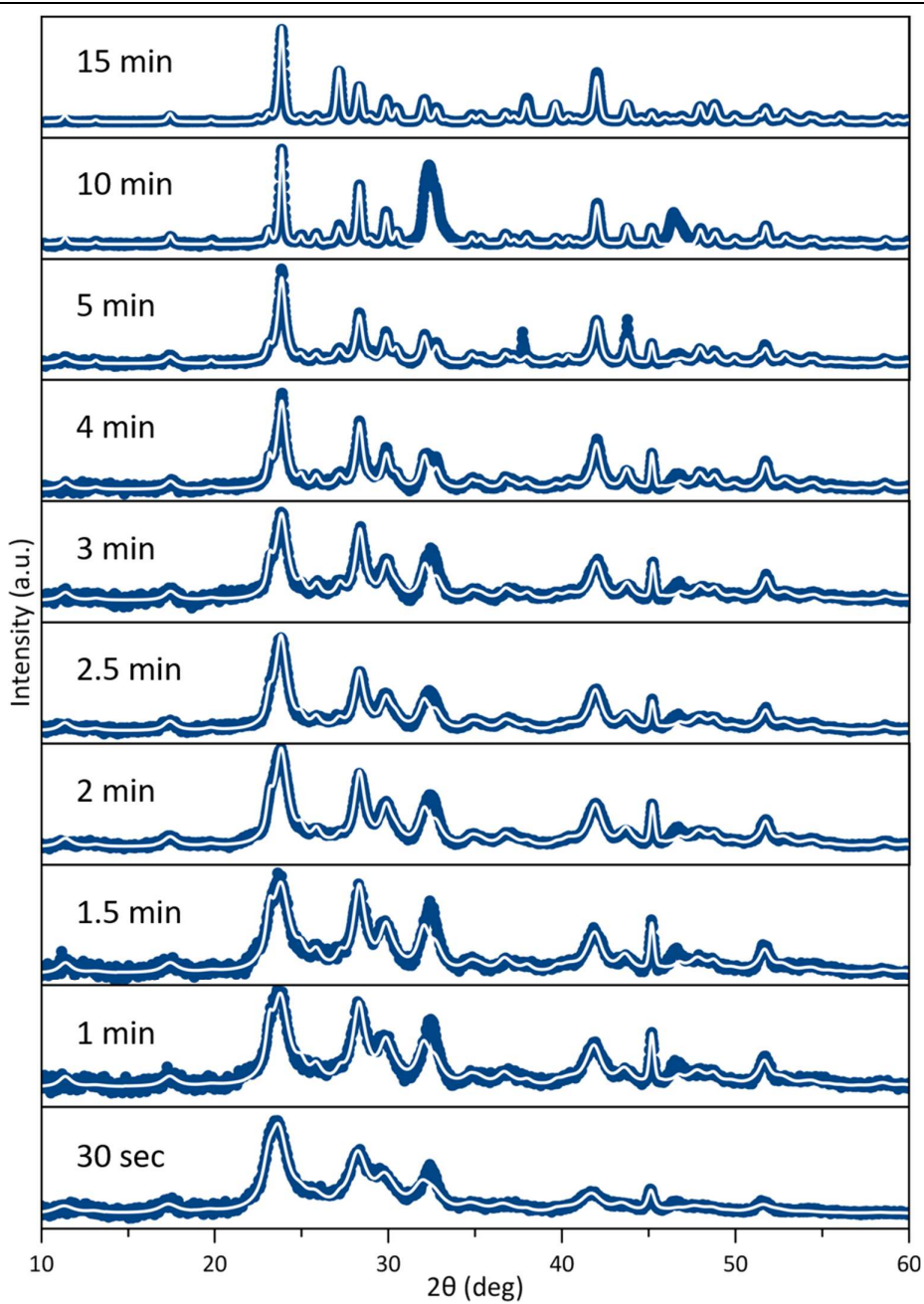

**Figure S9. XRD Rietveld Fits.** Rietveld fits of the XRD patterns collected from different reaction batch aliquots at various reaction times. Experimental patterns are shown in blue, while the fit traces are represented in white. To accurately capture the average dimensions of crystallites, we opted to exclude certain peaks (32.4° and 46.7°) from the fitting process when their significant preferential orientation and anisotropic broadening made it challenging to model their profiles accurately. This approach ensured that the fitting model reliably represented the remaining portions of the patterns.

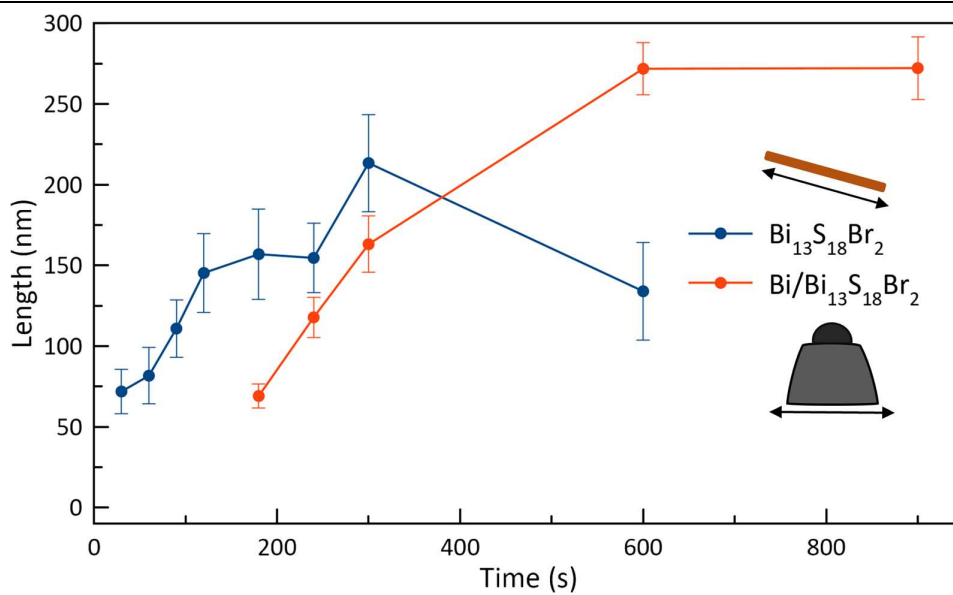

**Figure S10. Size evolution of Bi$_{13}$S$_{18}$Br$_2$ by TEM.** The length of free-standing Bi$_{13}$S$_{18}$Br$_2$ nanorods and of the Bi$_{13}$S$_{18}$Br$_2$ domains in Bi/Bi$_{13}$S$_{18}$Br$_2$ heterostructures was tracked by measuring 30 particles per time aliquot. Note note that there were not enough Bi/ Bi$_{13}$S$_{18}$Br$_2$ heterostructures to obtain reliable statistics before the 200-seconds mark.

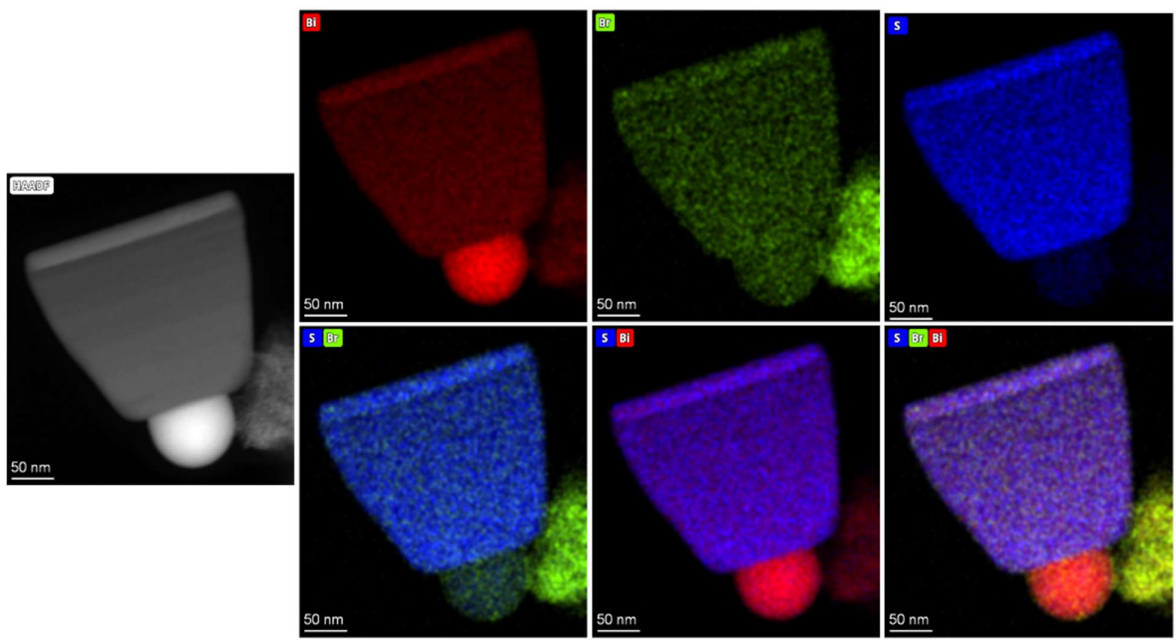

**Figure S11:** HAADF image and STEM-EDX compositional map of an individual nanobell, showing the presence of Bi, S and Br atoms.

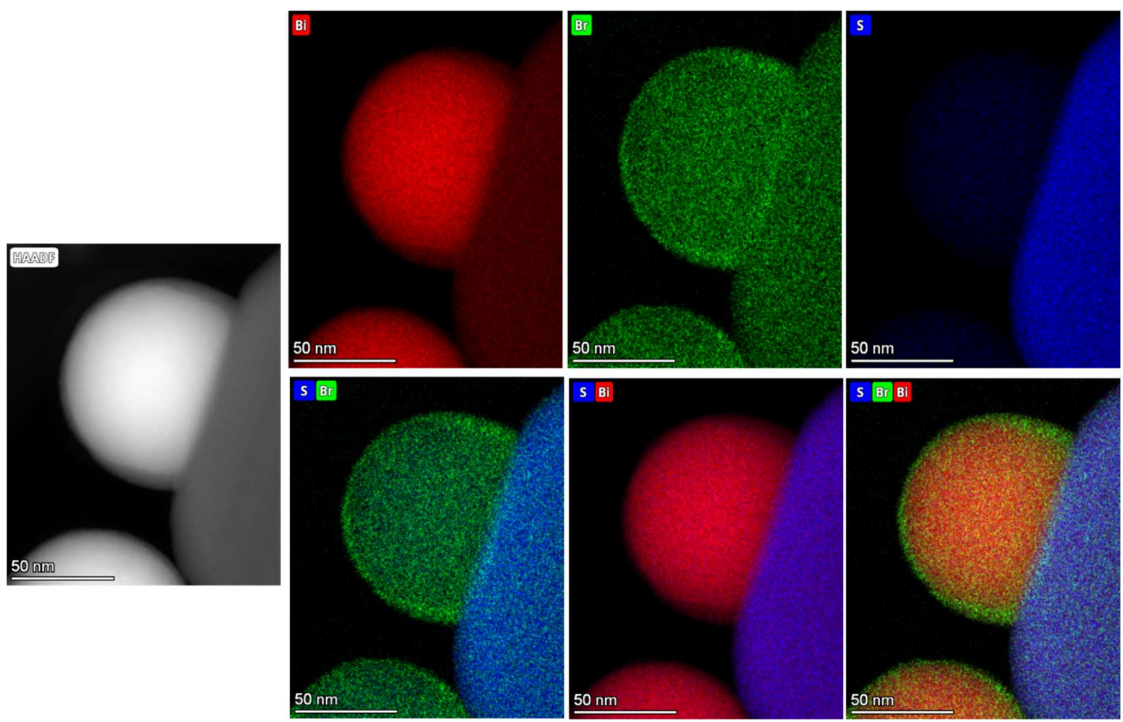

**Figure S12**: HAADF image and STEM-EDX compositional map of the Bi/chalcohalide contact region, showing the presence of a Br-rich shell surrounding the metal hemisphere.

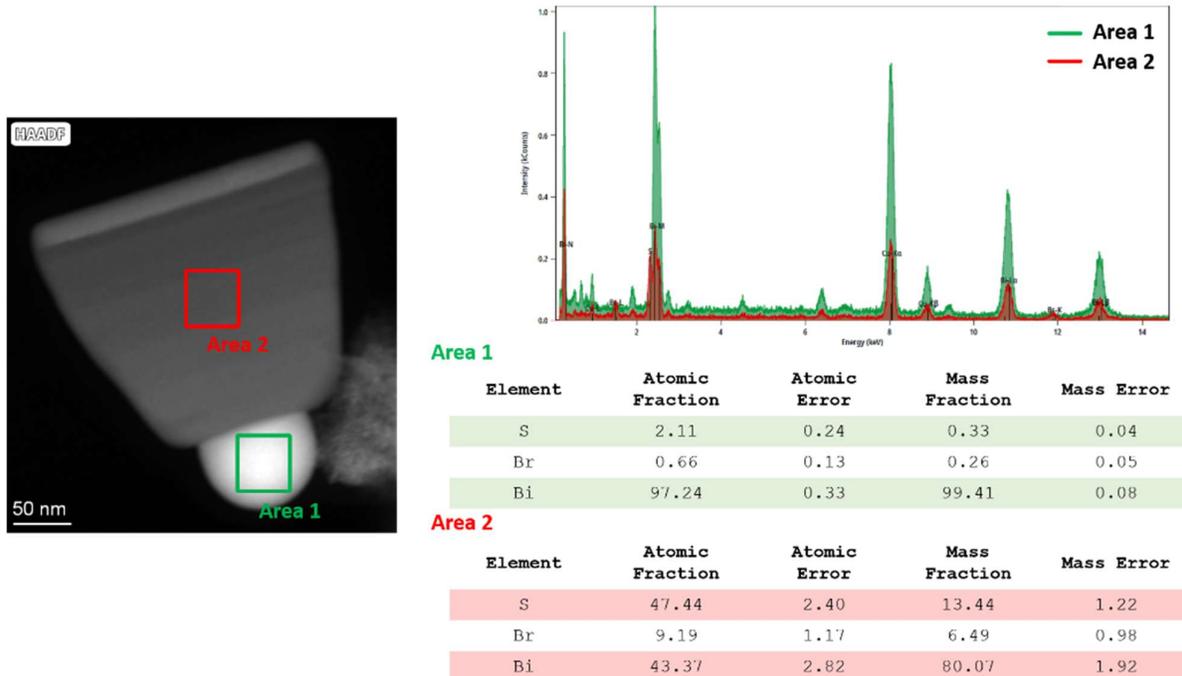

**Figure S13:** HAADF image of an individual nanobell and Energy-dispersive X-ray (EDX) analysis.

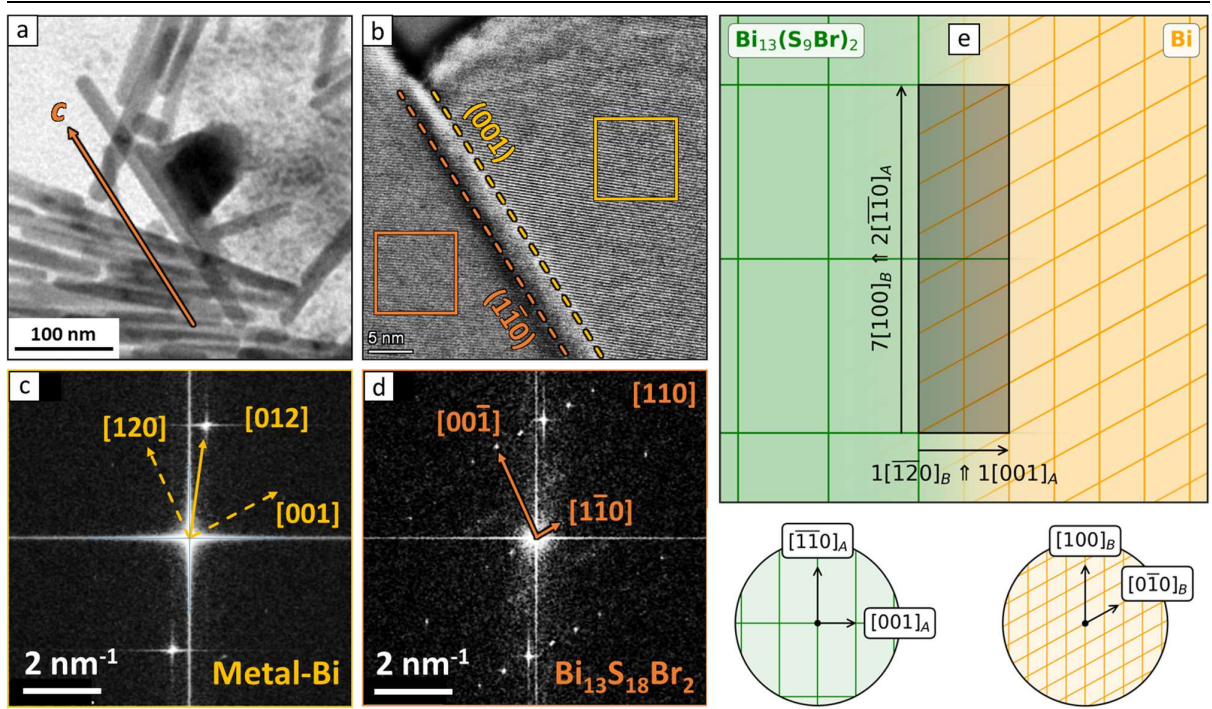

**Figure S14. Identification of the 001//1-10 – Bi/Bi$_{13}$S$_{18}$Br$_2$ epitaxial relation.** a) The morphology of early-stage heterostructures suggests that metallic Bi attaches to the sides of chalcohalide rods, which grow preferentially along the [001] direction. This implies that Bi$_{13}$S$_{18}$Br$_2$ exposes a (hk0) plane at the interface. b-d) Fourier transforms of HAADF images (b) support this hypothesis, as indicated by the three parallel stripes of reflections corresponding to the [hk-1], [hk0], and [hk1] planes in reciprocal space (d). The Bi orientation, however, could not be fully reconstructed from FFT due to the visibility of only a single reflection, identified as [012] based on its real-space periodicity (d = 3.28 Å). To resolve this, we used the Ogre library's lattice matching algorithm[65] to screen potential (hkl)//(1-10) Bi/Bi$_{13}$S$_{18}$Br$_2$ interfaces (with Bi-h, k, l < 2). The epitaxial relation was evaluated by its ability to reproduce the observed relative orientation of the lattice vectors [001]-Bi$_{13}$S$_{18}$Br$_2$ and [012]-Bi. The analysis identified the (001)//(1-10) Bi/Bi$_{13}$S$_{18}$Br$_2$ relation as the most suitable, characterized by a 2D supercell with an area of 248 Å² and 2.4% strain, as illustrated in panel (e). The dashed lines in panel (c) mark the orientations of the Bismuth lattice vectors as predicted by Ogre. For details on the lattice matching procedure and supercell interpretation, see Ref.II.

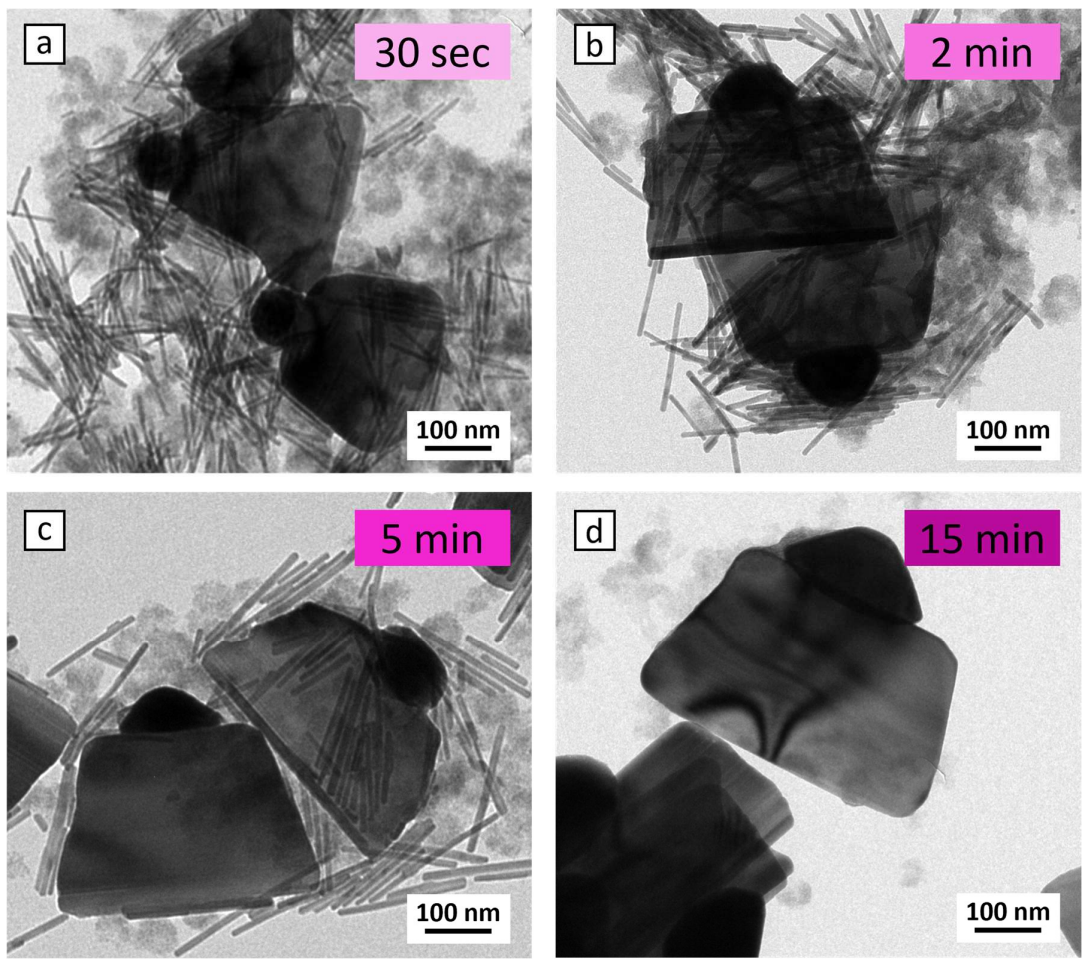

**Figure 15. Ostwald ripening growth test.** TEM images of Bi/Bi$_{13}$S$_{18}$Br$_2$ solution aliquots taken at a) 30 sec, b) 2 min, c) 5 min and d) 15 min after the two crude reaction solutions (one of nanorods and one of bells) have been put to react together at 180°C. The morphology evolution highlights the transfer of material from the nanorods to the nano-bells, which confirms the hypothesis of an Ostwald ripening-mediated growth mechanism. Interestingly, it is also evident how the bismuth domain becomes progressively faceted, and how at some point the chalcohalide resumes expanding along the *c*-preferred orientation growth tipical of the Bi$_{13}$S$_{18}$Br$_2$ phase.

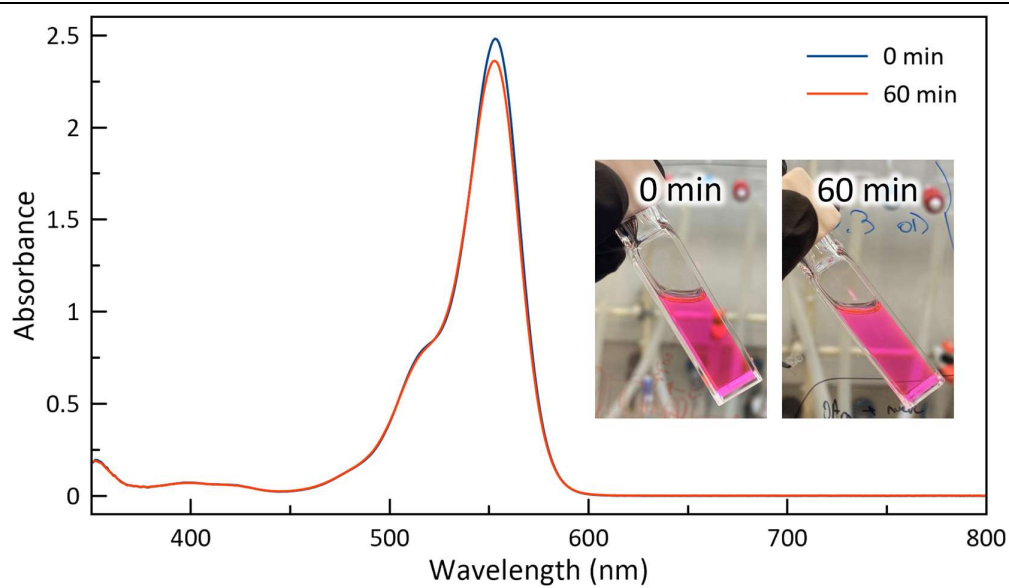

**Figure S16. Rhodamine-B photostability test in the absence of photocatalyst.** Absorbance spectra of the RhB solution before and after 1h of illumination at 420 nm in the absence of the photocatalyst.

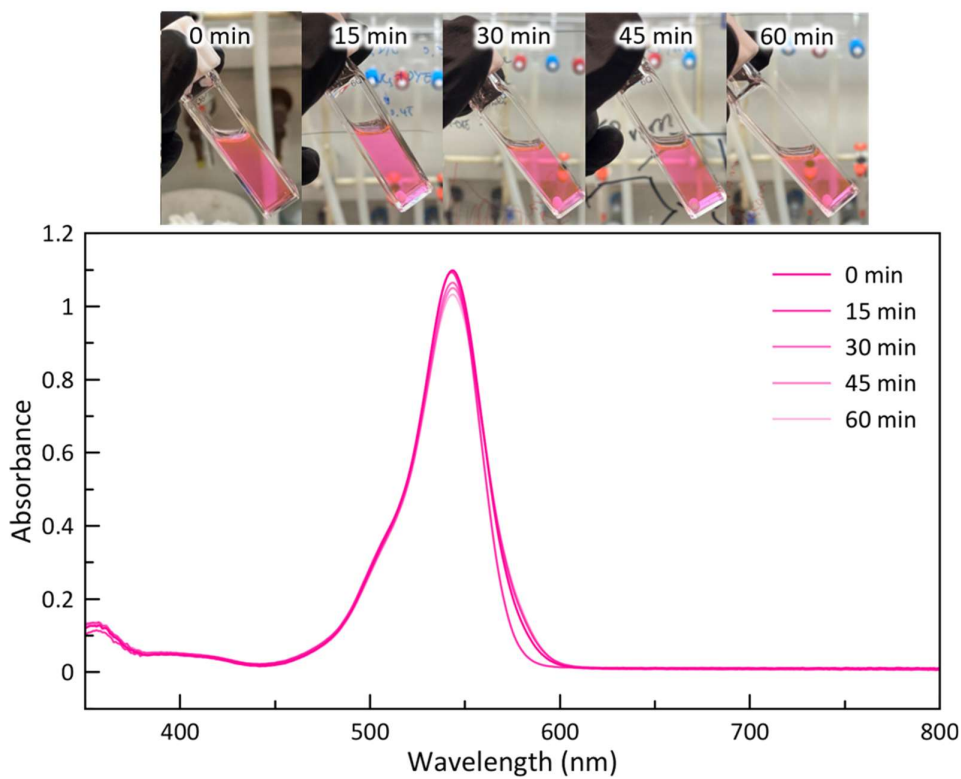

**Figure S17. Rhodamine-B test in dark condition.** Cuvettes images during the reaction time in dark condition and absorbance spectra of the RhB during the reaction time in dark condition, the photocatalyst was removed by centrifugation.

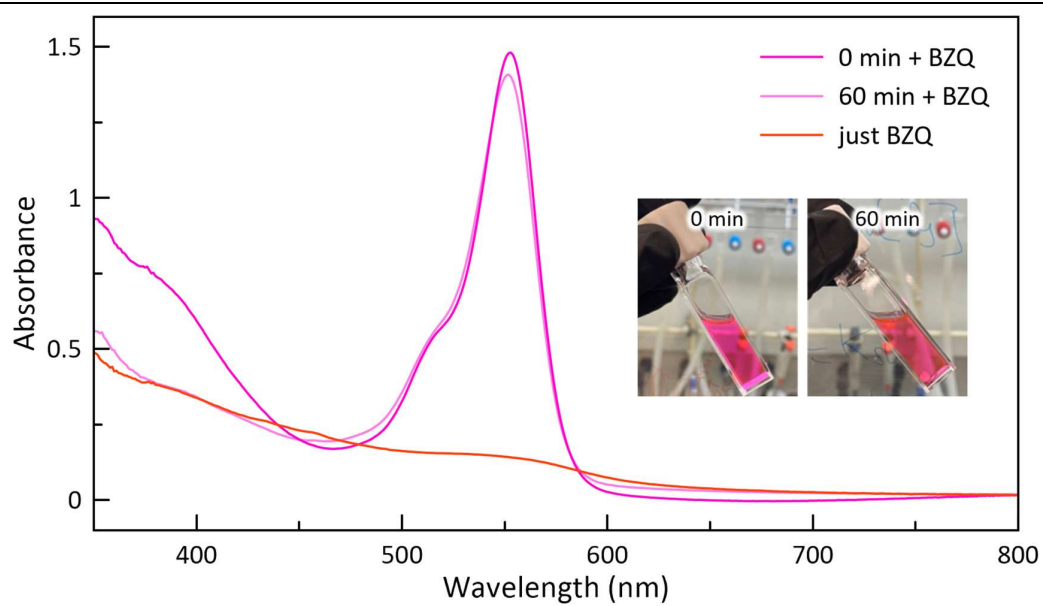

**Figure S18. Rhodamine-B test in the presence of benzoquinone (BZQ).** Cuvettes pictures before and after illumination with benzoquinone and absorbance spectra of RhB before and after illumination in the presence of benzoquinone, the photocatalyst was removed by centrifugation.

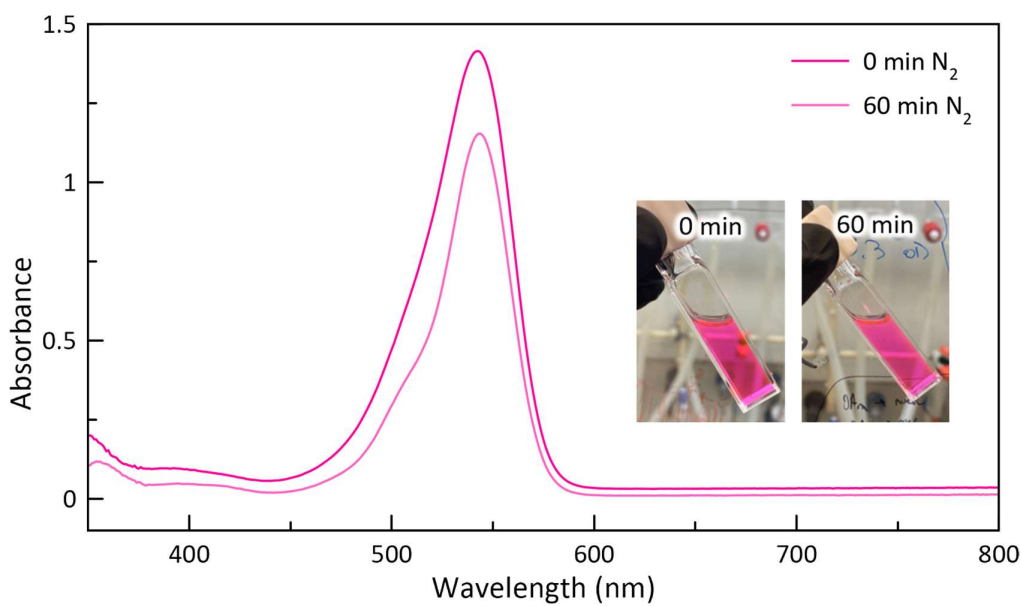

**Figure S19. Rhodamine-B test in inert atmosphere (N$_2$).** Cuvettes pictures before and after illumination in N$_2$ and absorbance spectra of RhB before and after illumination in N$_2$, the photocatalyst was removed by centrifugation.

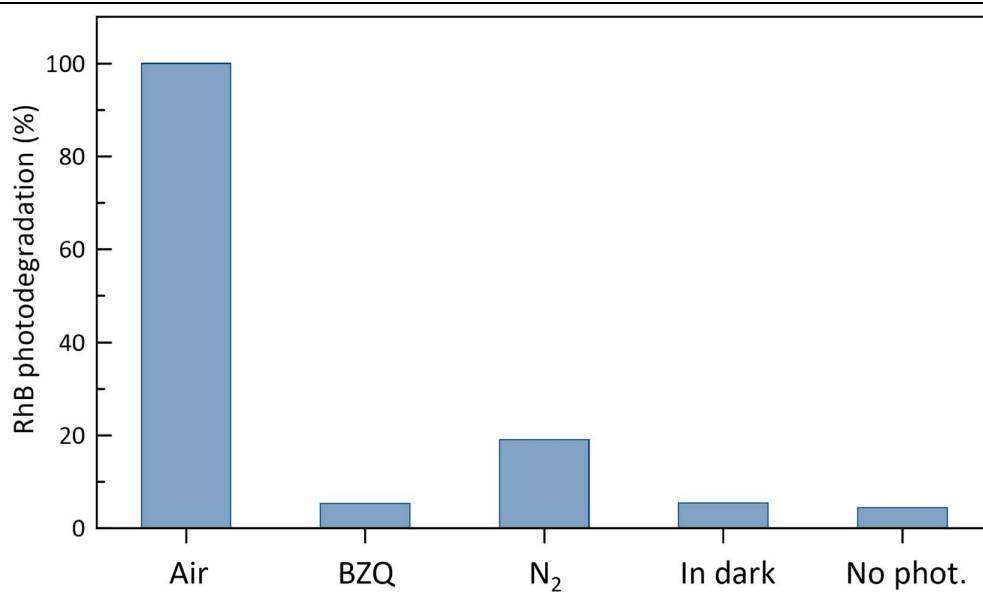

**Figure S20. RhB photodegradation percentage in different conditions.** Photodegradation activity of Bi/ Bi$_{13}$S$_{18}$Br$_2$ HSs in air condition, in the presence of benzoquinone (BZQ), in nitrogen atmosphere (N$_2$), in dark condition and in the absence of the photocatalyst (no phot.)

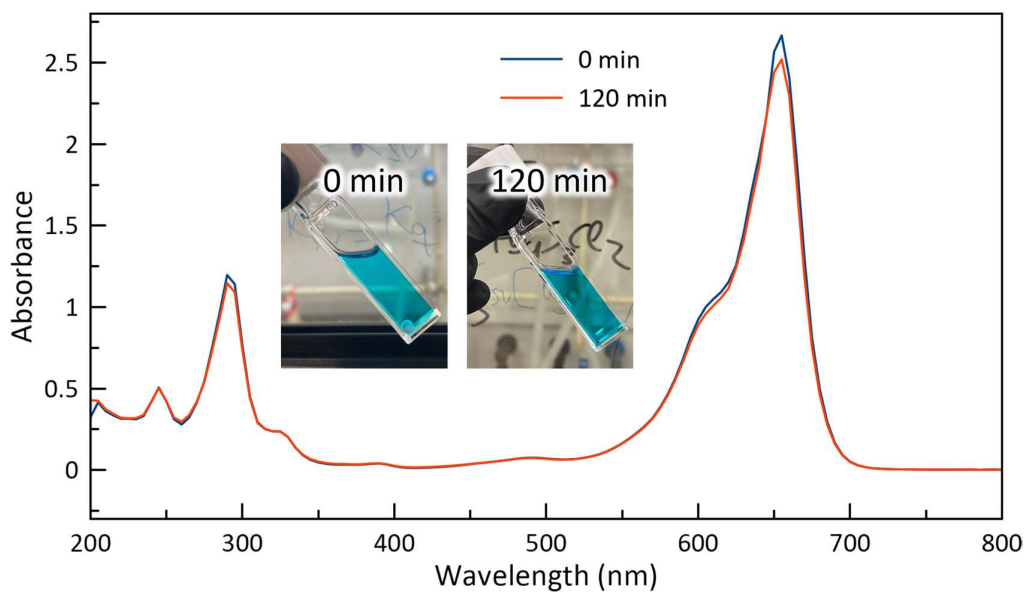

**Figure S21. Methylene Blue photostability test in the absence of photocatalyst.** Absorbance spectra of the MB solution before and after 2h of illumination at 420 nm in the absence of the photocatalyst.

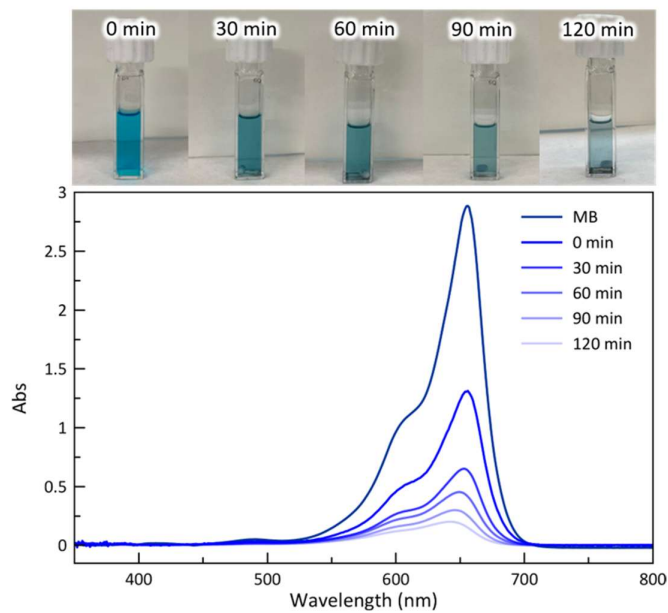

**Figure S22. Methylene Blue photodegradation process.** Cuvettes images during the reaction time in light condition and absorbance spectra of MB during the reaction time in light condition, the photocatalyst was removed by centrifugation.

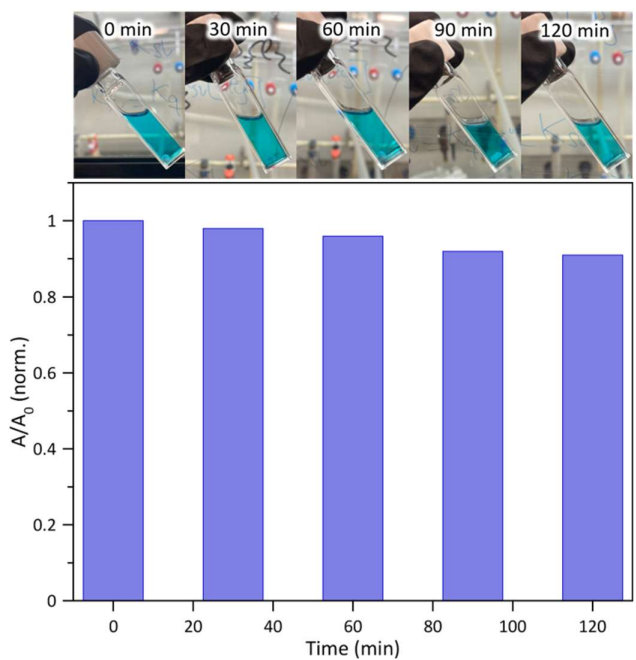

**Figure S23. Methylene Blue test in dark condition.** Cuvettes images during the reaction time in dark condition and normalized absorbance of MB during the reaction time in dark condition, the photocatalyst was removed by centrifugation.

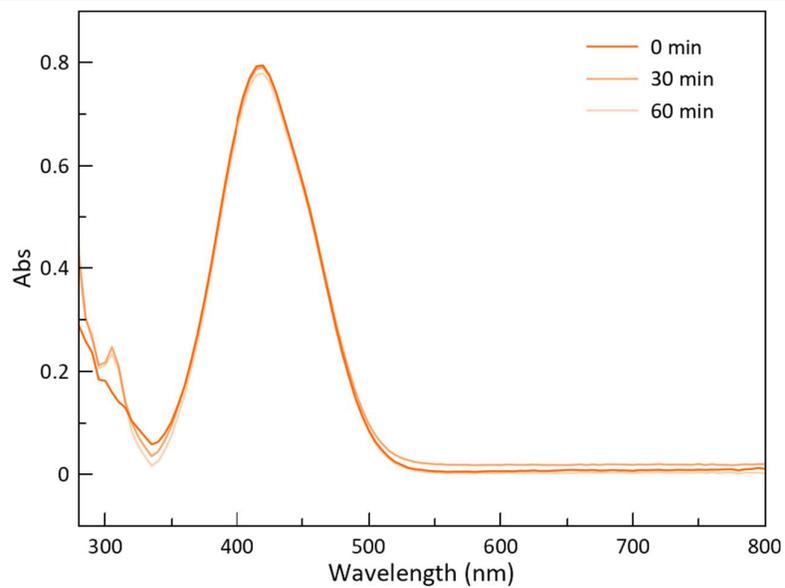

**Figure S24. Methyl Orange photodegradation process.** Absorbance spectra of MO during the reaction time in light condition, the photocatalyst was removed by centrifugation.

**Table S1. Z potential of Bi/Bi$_{13}$S$_{18}$Br$_2$.**

| Name | Mean |
|---|---|
| Zeta Potential (mV) | -58,299 |
| Conductivity (mS/cm) | 0 |
| Wall Zeta Potential (mV) | 0 |
| Quality Factor | 1,037 |
| Zeta Peak 1 Mean (mV) | -53,796 |

Open Circuit Voltage measures

Figure S25 gathers the OCP traces collected on control (a) and Bi HSs (b) samples. Upon illumination with a 415 nm light source, the open circuit voltage of control sample and heterostructures show a decrease in absolute value (i.e., the potential shifts towards more cathodic values). In both cases, this behavior is transient and reversible. Noteworthy, Bi HSs exhibit the noisier and complex trend (Figure S25b), with several spikes in the OCP value during the illumination. Nonetheless, a clear decrease/increase when the illumination is switched on/off is evident throughout the whole graph, thus confirming the interaction between Bi HSs and the light source.

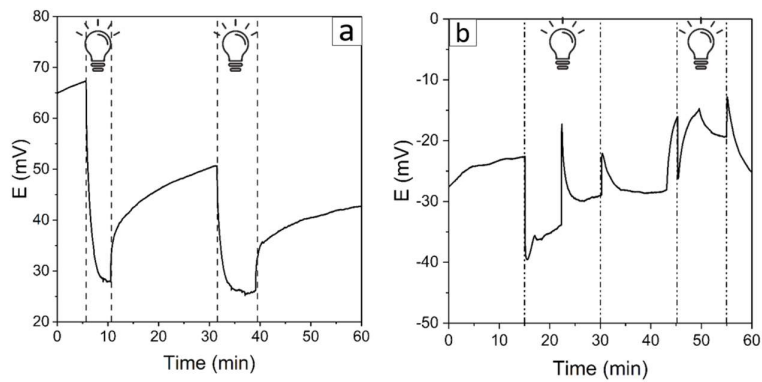

**Figure S25. Open Circuit Voltage (OCP) traces**. OCP traces recorded on (a) bare FTO, (b) Bi/ $Bi_{13}S_{18}Br_2$ HSs upon intermittent illumination with a 415 nm light source.

Cyclic Voltammetry (CV) of control samples

The CV collected on the bare support is gathered in Figure S26. As discussed in the main text, CVs do not show a significant variation upon illumination. This might be due to the limited contribution of the photogenerated currents with respect with the currents generated by the direct voltage application.

As for Bi HSs electrodes, CVs of bare FTO show additional voltammetric features when $CO_2$ is fed to the system, indicating a possible interaction of the material with $CO_2$. However, it must be noticed that the potentials screened in the bare FTO testing are way more cathodic than those investigated for Bi-based samples (Figure S25). This indicates a general inertness of FTO towards both HER and $CO_2$RR.

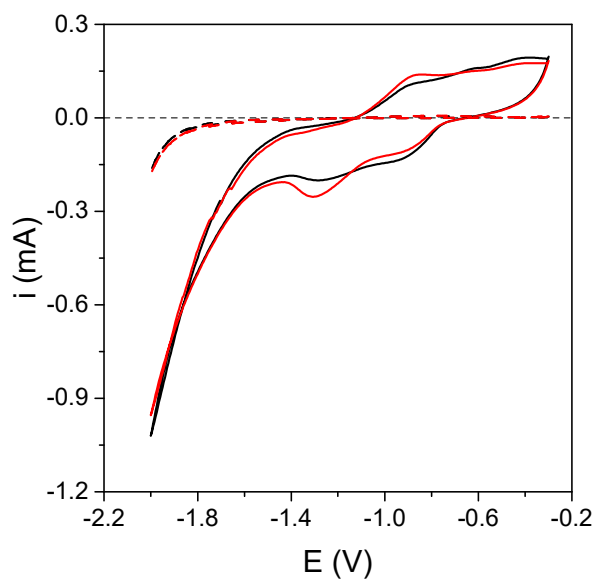

**Figure S26. Cyclic Voltammetry (CV) traces.** CV traces collected on bare FTO. (v = 100 mV s-1). Dotted curves have been collected under Ar bubbling (blank tests, only HER is possible), while full traces are related to CO2RR tests. Black: dark conditions. Red: illumination (415 nm LED). (b) CA scan at -1 V vs RE under intermittent illumination.

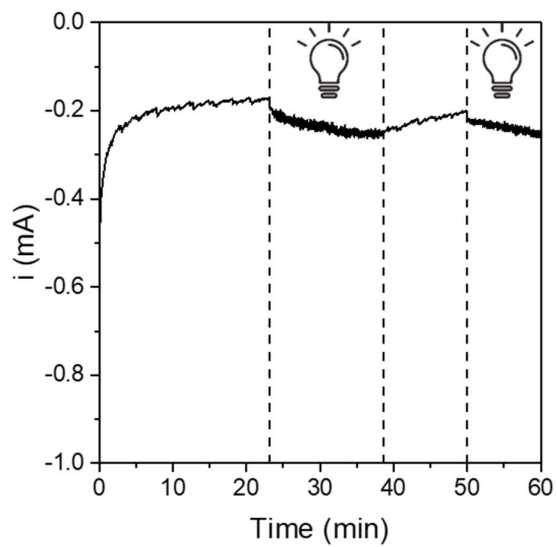

**Figure S27: Chronoamperometry (CA).** CA scans collected on bare FTO (@ -2 V vs RE) under $CO_2$RR conditions.

Gas and liquid-phase products detection

Despite the photoelectrochemical tests (especially those registered on Bi HSs electrodes) consistently indicate $CO_2$RR activity, the detection and quantification of products (also parasitic $H_2$, from HER) is hampered by the low currents delivered (i.e., low production rates). Indeed, the adaptation of our electrochemical $CO_2$RR setups, typically operated under flowing conditions, does not allow for the concentration of gas phase products in a static cell headspace, resulting in diluted samples, with concentrations lower than the instrument detection limit (Figure S28a). On the other hand, despite the expected concentration of liquid-phase $CO_2$RR products in the electrolyte, no products could be reliably quantified from HPLC chromatographs (Figure S28b).

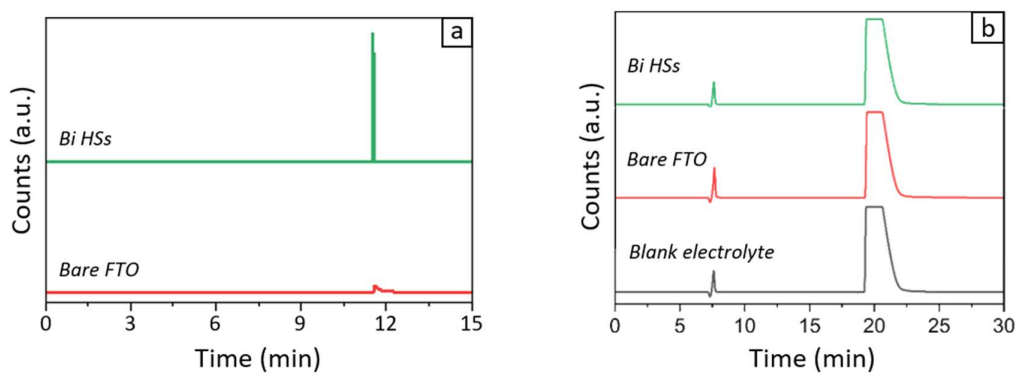

**Figure S28. Gas chromatography and liquid chromatography.** Sample chromatographic traces collected on the outlet gases and post-reaction electrolytes collected upon photoelectrochemical testing. (a) Typical FID (GC) and (b) RID (HPLC) traces.